\documentclass[12pt]{article}
\usepackage[left=1in,top=1in,right=1in,bottom=1in,head=.1in,nofoot]{geometry}
\usepackage{setspace,url,bm,amsmath} 

\usepackage{titlesec} 
\titlelabel{\thetitle.\quad} 
\titleformat*{\section}{\bf\large\center\uppercase} 

\usepackage[round]{natbib}
\usepackage{amssymb}
\usepackage{amsmath}
\usepackage{MnSymbol}
\usepackage{graphicx}
\usepackage{float}
\usepackage{ragged2e}
\usepackage{setspace}
\usepackage{blindtext}
\usepackage{bbm}
\usepackage{latexsym}
\usepackage{caption}
\usepackage{hyperref}
\usepackage{float}
\usepackage{comment}
\usepackage{dcolumn}
\usepackage{adjustbox}
\usepackage{xcolor}
\usepackage{setspace}
\usepackage{authblk}
\usepackage{subcaption}

\newcolumntype{Y}{>{\centering\arraybackslash}X}

\title{Randomization Tests that Condition on Non-Categorical Covariate Balance}
\author[1]{Zach Branson\thanks{
    We would like to thank two anonymous reviewers and the Associate Editor, Peter Aronow, for their insightful comments that led to notable improvements in this work. This research was supported by the National Science Foundation Graduate Research Fellowship Program under Grant No. 1144152. Any opinions, findings, and conclusions or recommendations expressed in this material are those of the authors and do not necessarily reflect the views of the National Science Foundation.}}
\author[2]{Luke Miratrix}

\affil[1]{Department of Statistics, Harvard University}
\affil[2]{Graduate School of Education and Department of Statistics, Harvard University}

\date{}

\begin{document}

\maketitle

\noindent
A benefit of randomized experiments is that covariate distributions of treatment and control groups are balanced on average, resulting in simple unbiased estimators for treatment effects. However, it is possible that a particular randomization yields covariate imbalances that researchers want to address in the analysis stage through adjustment or other methods. Here we present a randomization test that conditions on covariate balance by only considering treatment assignments that are similar to the observed one in terms of covariate balance. Previous conditional randomization tests have only allowed for categorical covariates, while our randomization test allows for any type of covariate. Through extensive simulation studies, we find that our conditional randomization test is more powerful than unconditional randomization tests and other conditional tests. Furthermore, we find that our conditional randomization test is valid (1) unconditionally across levels of covariate balance, and (2) conditional on particular levels of covariate balance. Meanwhile, unconditional randomization tests are valid for (1) but not (2). Finally, we find that our conditional randomization test is similar to a randomization test that uses a model-adjusted test statistic.

\noindent
\section{After Randomization: To Adjust or Not To Adjust?}

Randomized experiments are often considered the ``gold standard'' of statistical inference because randomization balances the covariate distributions of the treatment and control groups on average, which limits confounding between treatment effects and covariate effects. However, it is possible that a particular treatment assignment from a randomized experiment yields covariate imbalances that researchers wish to address. One option is to employ experimental design strategies such as blocking or rerandomization \citep{morgan2012rerandomization}, which prevent substantial covariate imbalance from occurring before the experiment is conducted. 
When these strategies are not employed, covariate imbalance must be addressed in the analysis stage rather than the design stage. 
The analyst of such experiments must make a choice: to adjust or not to adjust for the covariate imbalance realized by a particular randomization. 
If adjustment is done, it is typically done via statistical models (e.g., regression adjustment); however, the results from such adjustment may be biased and/or sensitive to model specification \citep{imai2008misunderstandings,freedman2008regression,aronow2013class}. 
Meanwhile, unadjusted estimators---though unbiased across randomizations---could be confounded by the realized covariate imbalance at hand.
\cite{lin2013agnostic} rigorously investigated these tradeoffs between unadjusted and adjusted estimators, noting that biases due to regression are often minimal, but also that unadjusted estimators are appealing for their simplicity and transparency. 
Regardless of where these works fall on the ``to adjust or not to adjust'' spectrum, they all agree that accounting for covariate balance is a key concern in randomized experiments.

\subsection{Accounting for Covariate Balance in Randomization Tests}
 
In addition to model-based testing, one can use randomization tests to account for covariate balance in experiments. Randomization tests are often considered minimal-assumption approaches in that they usually only require assuming a probability distribution on treatment assignment rather than structural modeling assumptions or central limit theorems \citep{rosenbaum2002observational}. 
In particular, a randomization test requires specifying only (1) the assumed assignment mechanism and (2) the test statistic. 
In this context, one can account for covariate balance by making particular choices for the assignment mechanism or the test statistic, but most have focused on the choice of the latter. 
For example, many have found that using model-adjusted estimators as test statistics to address covariate imbalances can result in statistically powerful randomization tests (\citealt{raz1990testing}, \citealt{rosenbaum2002covariance}, \citealt{rosenbaum2002observational} Chapter 2, \citealt{hernandez2004covariate}, \citealt{imbens2015causal} Chapter 5). 
Meanwhile, practitioners typically use the assignment mechanism that was actually used in the design of the experiment when conducting a randomization test (e.g., if units were assigned completely at random, then this same assignment mechanism is used during the randomization test). 
However, by considering other choices for the assignment mechanism, one can also account for covariate balance.

In particular, a small strand of literature has explored randomization tests that restrict the assignment mechanism to only consider treatment assignments that are similar to the observed one in terms of covariate balance, even if such an assignment mechanism was not explicitly specified by design. This literature has focused on cases where all covariates are categorical, and thus treatment assignment is characterized by permutations within covariate strata. For example, \cite{rosenbaum1984conditional} proposed a conditional permutation test for observational studies that permutes the treatment indicator within groups of units with the same covariate values. This test assumes (1) the treatment assignment is strongly ignorable, (2) the true propensity score model is a logistic regression model, and (3) the collection of covariates is sufficient for the logistic regression model. More recently, \cite{hennessy2016conditional} proposed a conditional randomization test for randomized experiments that is similar to \cite{rosenbaum1984conditional} in that it also permutes within groups of units with the same covariate values, but it does not require any kind of model specification. \cite{rosenbaum1984conditional} and \cite{hennessy2016conditional} only consider cases with categorical covariates, and they make connections between their randomization tests and adjustment methods for categorical covariates, such as post-stratification \citep{miratrix2013adjusting}.

\subsection{Our Contribution: Considering Non-Categorical Covariates}

We develop a randomization test that conditions on the realized covariate balance of an experiment for the more general case where covariates may be non-categorical. We demonstrate that our randomization test is more powerful than randomization tests that do not condition on covariate balance and is comparable to randomization tests that use model-adjusted estimators as test statistics. In general, we recommend the use of randomization tests that either condition on covariate balance through the assignment mechanism or utilize model-adjusted test statistics, instead of an unconditional randomization test that uses an unadjusted test statistic.

Our main contribution is outlining a randomization test that conditions on covariate balance through the assignment mechanism for the general case of non-categorical covariates. Unlike the case where only categorical covariates are present, samples from the conditional randomization distribution cannot be obtained via permutations of the treatment indicator when there are non-categorical covariates. In response to this complication, we develop a rejection-sampling algorithm to sample from the conditional randomization distribution.

We find that our conditional randomization test appears to be equivalent to randomization tests that use regression-based test statistics. 
This contribution is particularly notable because most have characterized the choice of test statistic as the main avenue for increasing the power of a randomization test and for adjusting for imbalance in an experiment. 
Our work suggests how the choice of assignment mechanism can be an analogous avenue for obtaining statistically powerful randomization tests that appropriately adjust for imbalance. 
Furthermore, through simulation, we also find that our conditional randomization test is valid across randomizations conditional on a particular level of covariate balance, while unconditional randomization tests are often not valid across such randomizations.
This suggests that our conditional randomization test can be used to ensure that statistical inferences are valid for the observed data at hand; meanwhile, unconditional randomization tests do not provide this benefit.
Overall this suggests that practitioners using randomization tests should either condition on observed imbalance or use adjusted test statistics rather than the traditional randomization procedures usually seen in the literature.

To build intuition for our conditional randomization test, in Section \ref{s:reviewRandomizationTests} we review randomization tests for Fisher's Sharp Null and review the conditional randomization test of \cite{hennessy2016conditional}. In Section \ref{s:conditionalRandomizationTest} we outline our conditional randomization test, which can flexibly condition on multiple levels of balance for non-categorical covariates. In Section \ref{s:simulations} we provide simulation evidence that our conditional randomization test (1) is more powerful than unconditional and other conditional randomization tests, and (2) is approximately equivalent to an unconditional randomization test that uses a regression-based test statistic. In Section \ref{s:discussionAndConclusion} we conclude by discussing how confidence intervals can be constructed from our conditional randomization test and the extent to which our conditional randomization test can be used for observational studies.

\section{Review of Randomization Tests for Fisher's Sharp Null} \label{s:reviewRandomizationTests}

We focus on randomization tests for Fisher's Sharp Null. While conclusions from such tests are limited---the only conclusion that can be made is whether or not there is any treatment effect among the experimental units---in Section \ref{s:discussionAndConclusion} we discuss how such tests can be inverted to yield uncertainty intervals as well.

First we review a general framework for randomization tests for Fisher's Sharp Null. We then review the unconditional randomization test typically discussed in the literature under this framework. Finally, we review the conditional randomization test of \cite{hennessy2016conditional} that conditions on categorical covariate balance. 

\subsection{Setup and Randomization Test Procedure} \label{ss:setupProcedureRandomizationTests}

Consider $N$ units to be allocated to treatment and control in a randomized experiment. Following \cite{rubin1974estimating}, let $Y_i(1)$ and $Y_i(0)$ denote the treatment and control potential outcomes, respectively, for unit $i = 1,\dots,N$, and let $\mathbf{x}_i$ denote a $p$-dimensional vector of pre-treatment covariates. Let $W_i = 1$ if unit $i$ is assigned to treatment and 0 otherwise. Furthermore, define $\mathbf{X} \equiv (\mathbf{x}_1,\dots,\mathbf{x}_N)^T$ and $\mathbf{W} \equiv (W_1,\dots,W_N)$ as the covariate matrix and vector of treatment assignments, respectively. The observed outcomes are
$y_i  = W_iY_i(1) + (1-W_i)Y_i(0)$.
Importantly, the potential outcomes $(Y_i(1), Y_i(0))$ and covariates $\mathbf{x}_i$ are fixed; the only stochastic element of the observed outcomes $y_i$ is the treatment assignment $W_i$.

Throughout, we assume a completely randomized experiment, where the true distribution of the treatment assignment $\mathbf{W}$ is:
	\begin{align}
		P(\mathbf{W} = \mathbf{w}) = \begin{cases}
			{N \choose N_T}^{-1} &\mbox{ if } \sum_{i=1}^N w_i = N_T \\
			0 &\mbox{ otherwise.}
		\end{cases} \label{eqn:completeRandomization}
	\end{align}
with the number of treated units, $N_T$, fixed. Many causal estimands can be considered in this framework, but we focus on the average treatment effect
\begin{align}
	\tau = \frac{1}{N} \sum_{i=1}^N \big(Y_i(1) - Y_i(0) \big) \label{eqn:tau}
\end{align}
because it is the most common estimand in the causal inference literature. The potential outcomes $Y_i(1)$ and $Y_i(0)$ are never both observed, so (\ref{eqn:tau}) needs to be estimated. One common estimator is the mean-difference estimator
\begin{align}
	\hat{\tau}_{sd} = \frac{\sum_{i=1}^{N} W_i Y_i(1)}{\sum_{i=1}^N W_i} - \frac{\sum_{i=1}^N (1-W_i)Y_i(0)}{\sum_{i=1}^N (1 - W_i)} = \frac{\sum_{i: W_i = 1} y_i}{N_T} - \frac{\sum_{i: W_i = 0} y_i}{N_C} = \bar{y}_T - \bar{y}_C \label{eqn:meanDiffEst}
\end{align}
where $N_T \equiv \sum_{i=1}^N W_i$ and $N_C \equiv \sum_{i=1}^N (1 - W_i)$ are the number of units that receive treatment and control, respectively.

A common test for assessing if an estimate for the average treatment effect is statistically significant is to test for Fisher's Sharp Null:
\begin{align}
	H_0: Y_i(1) = Y_i(0), \hspace{0.1 in} \forall i = 1,\dots, N
\end{align}
which states that there is no treatment effect for any of the $N$ units. A rejection of Fisher's Sharp Null implies that a treatment effect is present. We focus on testing Fisher's Sharp Null because it is the most common hypothesis to assess using randomization tests in the causal inference literature \citep{rosenbaum2002observational,imbens2015causal}. See \cite{ding2017paradox} and the ensuing comments for a discussion of how testing Fisher's Sharp Null compares to testing Neyman's Weak Null within the context of randomization-based causal inference.

Under Fisher's Sharp Null, the outcomes for any particular randomization will be equal to the observed outcomes; i.e., the observed outcomes will be the same across all realizations of $\mathbf{W}$ under the Sharp Null. Thus, under $H_0$, the value of any test statistic $t\big(Y(\mathbf{W}), \mathbf{W}, \mathbf{X} \big)$ can be computed for any particular realization of the treatment assignment $\mathbf{W}$. A common choice of test statistic is $t\big(Y(\mathbf{W}), \mathbf{W}, \mathbf{X} \big) = \hat{\tau}_{sd}$. Our framework can incorporate any test statistic that differentiates between treatment and control response; for now we will focus on the test statistic $\hat{\tau}_{sd}$, and later we will discuss model-adjusted test statistics. See \citet[Chapter 2]{rosenbaum2002observational} for further discussion on choices of test statistics for randomization tests.

To test Fisher's Sharp Null, one compares the observed value of the test statistic, $t^{obs}$, to the randomization distribution of the test statistic under the Sharp Null. Importantly, the randomization distribution of the test statistic depends on the set of treatment assignments that one considers possible within the randomization test.

We follow the notation of \citet[Chapter 4]{imbens2015causal} and define $\mathbb{W}$ as the set of treatment assignments with positive probability within a given randomization test. Given any test statistic $t\big(Y(\mathbf{W}), \mathbf{W}, \mathbf{X} \big)$, the two-sided randomization test $p$-value for Fisher's Sharp Null is
\begin{align}
    P \big( \big|t \big(Y(\mathbf{W}), \mathbf{W}, \mathbf{X} \big) \big| \geq |t^{obs}| \big) &= \sum_{\mathbf{w} \in \mathbb{W}} \mathbb{I} \big( \big|t \big(Y(\mathbf{w}), \mathbf{w}, \mathbf{X} \big) \big| \geq |t^{obs}| \big)P(\mathbf{W} = \mathbf{w}) \label{randomizationTestPValue}
\end{align}
In other words, the $p$-value (\ref{randomizationTestPValue}) is the probability that a test statistic larger than the observed one would have occurred under the Sharp Null, given the assignment mechanism $P(\mathbf{W})$.

Typically, the set $\mathbb{W}$ is too large to feasibly compute (\ref{randomizationTestPValue}). Instead, (\ref{randomizationTestPValue}) can be approximated by randomly sampling $\mathbf{w}^{(1)}, \dots, \mathbf{w}^{(M)}$ from $P(\mathbf{W})$; then, the randomization-test $p$-value (\ref{randomizationTestPValue}) is approximated by
\begin{align}
    P \big( \big|t \big(Y(\mathbf{W}), \mathbf{W}, \mathbf{X} \big) \big| \geq t^{obs} \big) &\approx \frac{ \sum_{m=1}^M \mathbb{I} \big( \big|t \big(Y(\mathbf{w}^{(m)}), \mathbf{w}^{(m)}, \mathbf{X} \big) \big| \geq \big|t^{obs}\big| \big)}{M} \label{randomizationTestPValueApproximationEqualProbabilites}
\end{align}
Thus, testing Fisher's Sharp Null is a three-step procedure \citep{good2013permutation}:
\begin{enumerate}
    \item Specify the distribution $P(\mathbf{W})$ (and, consequentially, $\mathbb{W}$) to be used within the randomization test.
    \item Choose a test statistic $t\big(Y(\mathbf{W}), \mathbf{W}, \mathbf{X} \big)$.
    \item Compute or approximate the $p$-value (\ref{randomizationTestPValue}).
\end{enumerate}
In the remainder of this section we will discuss two randomization tests: one that does not condition on covariate balance and one that does. The only difference between the two tests is the first step in the procedure above, i.e., the choice of the assignment mechanism $P(\mathbf{W})$.

\subsection{Unconditional Randomization Tests} \label{ss:unconditionalRandomizationTest}

The most common randomization test in the literature utilizes the same assignment mechanism used to design the experiment, the completely randomized assignment mechanism defined in (\ref{eqn:completeRandomization}). A completely randomized assignment mechanism assumes that $\mathbb{W} = \{ \mathbf{w} : \sum_{i=1}^N w_i = N_T\}$, i.e., it only considers assignments where $N_T$ units are assigned to treatment. \cite{hennessy2016conditional} call randomization tests that assume a completely randomized assignment mechanism ``unconditional randomization tests'' because they do not condition on forms of covariate balance. Once $P(\mathbf{W})$ and a test statistic are specified, the randomization test follows the three-step procedure from Section \ref{ss:setupProcedureRandomizationTests}. This test is also called a permutation test because random samples from $P(\mathbf{W})$ can be obtained by randomly permuting the observed treatment assignment $\mathbf{W}^{obs}$.

Instead of using $P(\mathbf{W})$ in the randomization test procedure, \cite{hennessy2016conditional} proposed using an assignment mechanism that conditions on covariate balance.

\subsection{Conditional Randomization Tests} \label{ss:conditionalRandomizationTest}

Because the number of treated units is prespecified as part of the design of a completely randomized experiment, the unconditional randomization test in Section \ref{ss:unconditionalRandomizationTest} follows the typical recommendation to ``analyze as you randomize.'' However, many have recommended conditioning on the observed number of treated units even when the number of treated units was not specified by design \citep{hansen2008covariate,zheng2008multi,miratrix2013adjusting,rosenberger2015randomization}. The goal of conditional inference in general (and conditional randomization tests specifically) is to focus inference on experiments that are most relevant to the data at hand by conditioning on pertinent statistics such as the number of treated units or forms of covariate balance. As we show through simulation in Section \ref{s:simulations}, conditional randomization tests can have the benefit of being valid conditional on the data as well as being valid unconditionally, whereas unconditional randomization tests are only valid unconditionally.

To formalize this idea of conditioning on pertinent statistics, define a criterion that is a function of the treatment assignment and pre-treatment covariates:
\begin{align}
    \phi(\mathbf{W}, \mathbf{X}) = \begin{cases}
        1 &\mbox{ if } \mathbf{W} \text{ is an acceptable treatment assignment} \\
        0 &\mbox{ if } \mathbf{W} \text{ is not an acceptable treatment assignment.}
    \end{cases}
\end{align}
This notation mimics that of \cite{morgan2012rerandomization}, who use $\phi(\mathbf{W}, \mathbf{X})$ to define treatment assignments that are desirable for an experimental design, and that of \cite{branson2018randomization}, who were the first to introduce such notation for randomization tests. The unconditional randomization test in Section \ref{ss:unconditionalRandomizationTest} inherently defines $\phi(\mathbf{W}, \mathbf{X}) = 1$ if $\sum_{i=1}^N W_i = N_T$ and 0 otherwise. In general, conditional randomization tests involve sampling from the conditional distribution $P(\mathbf{W} | \phi(\mathbf{W}, \mathbf{X}) = 1)$ rather than the unconditional distribution $P(\mathbf{W})$ in Section \ref{ss:unconditionalRandomizationTest}.

\cite{hennessy2016conditional} focus on $\phi(\mathbf{W}, \mathbf{X})$ that indicate some specified degree of categorical covariate balance. Assume there are covariate strata $s = 1, \dots , S$ specified by the researcher such that each unit belongs to only one stratum, and define $c_i = s$ if the $i$th unit belongs to the $s$th stratum. The strata may be defined using all of the covariates or some subset of them. Then, \cite{hennessy2016conditional} define the criterion $\phi(\mathbf{W}, \mathbf{X})$ as\footnote{\cite{hennessy2016conditional} use slightly different notation, instead defining a balance function $B(\mathbf{W}, \mathbf{X})$ and condition on the balance function being equal to some prespecified $b$. The more general notation that uses $\phi(\mathbf{W}, \mathbf{X})$ will become helpful in our discussion of continuous covariate balance.}
\begin{align}
    \phi_s(\mathbf{W}, \mathbf{X}) = \begin{cases}
        1 &\mbox{ if } \sum_{i: c_i = s} W_i = N_{T,s} \text{, for } s =1,\dots, S \\
        0 &\mbox{ otherwise.} \label{eqn:categoricalPhi}
    \end{cases}
\end{align}
In other words, each stratum is treated as a completely randomized experiment. \cite{hennessy2016conditional} assume that the conditional distribution $P(\mathbf{W} | \phi_s(\mathbf{W}, \mathbf{X}) = 1)$ is uniform, i.e.,
\begin{align}
	P(\mathbf{W} | \phi_s(\mathbf{W}, \mathbf{X}) = 1) = \begin{cases}
 		\left(\prod_{s=1}^S {N_s \choose N_{T,s}} \right)^{-1} &\mbox{ if } \sum_{i: c_i = s} W_i = N_{T, s} \text{, for } s = 1,\dots,S  \\
 		0 &\mbox{ otherwise.}
 	\end{cases} \label{eqn:strataRandomization}
\end{align}
Random samples from $P(\mathbf{W} | \phi_s(\mathbf{W}, \mathbf{X}) = 1)$ can be obtained by randomly permuting the observed treatment assignment $\mathbf{W}^{obs}$ within the covariate strata $s = 1,\dots,S$. Once a test statistic is specified, the conditional randomization test follows the three-step procedure in Section \ref{ss:setupProcedureRandomizationTests}, but using $P(\mathbf{W} | \phi_s(\mathbf{W}, \mathbf{X}) = 1)$ instead of $P(\mathbf{W})$.

\cite{hennessy2016conditional} showed via simulation that this conditional randomization test using the test statistic $\hat{\tau}_{sd}$ is more powerful than the unconditional randomization test in Section \ref{ss:unconditionalRandomizationTest} using $\hat{\tau}_{sd}$. Furthermore, they found that this conditional randomization test using $\hat{\tau}_{sd}$ is comparable to the unconditional randomization test using the post-stratification test statistic
\begin{align}
	\hat{\tau}_{ps} = \sum_{s=1}^S \frac{N_s}{N} \hat{\tau}_{sd}(s), \label{eqn:postStratEst}
\end{align}
where $\hat{\tau}_{sd}(s)$ is the estimator $\hat{\tau}_{sd}$ within stratum $s$ \citep{miratrix2013adjusting}.

Note that the set of possible treatment assignments $\mathbb{W}$ must be large enough to perform a powerful randomization test. For example, if $| \mathbb{W} | < 20$, then it is impossible to obtain a randomization test $p$-value less than 0.05. It may be surprising that conditional randomization tests can be more powerful than unconditional randomization tests, because the former utilizes fewer treatment assignments than the latter. However, these fewer treatment assignments are more relevant to the observed treatment assignment in terms of covariate balance, which leads to more powerful inference, as discussed by works such as \cite{rosenbaum1984conditional} and \cite{hennessy2016conditional}.

When the criterion $\phi(\mathbf{W}, \mathbf{X})$ is defined as in (\ref{eqn:categoricalPhi}), $| \mathbb{W} | = \prod_{s=1}^S {N_s \choose N_{T,s} }$, which is typically large. Furthermore, assuming that $P(\mathbf{W} | \phi(\mathbf{W}, \mathbf{X}) = 1)$ is uniform, random samples from this distribution can be obtained directly, and thus implementation of the conditional randomization test is straightforward. However, this approach is less straightforward when $\mathbf{X}$ contains non-categorical covariates, because $\mathbf{X}$ is no longer composed of strata where there are treatment and control units in each stratum. One option is to coarsen $\mathbf{X}$ into strata and then use the conditional randomization test of \cite{hennessy2016conditional}. Instead of throwing away information via coarsening, we propose a criterion $\phi(\mathbf{W}, \mathbf{X})$ that incorporates covariate balance for non-categorical covariates. We define $\phi(\mathbf{W}, \mathbf{X})$ such that $| \mathbb{W} |$ is large enough while still sufficiently conditioning on covariate balance. Furthermore, as we discuss below, random samples from $P(\mathbf{W} | \phi(\mathbf{W}, \mathbf{X}) = 1)$ will no longer be equivalent to random permutations of $\mathbf{W}^{obs}$; thus, we develop an algorithm to obtain random samples from $P(\mathbf{W} | \phi(\mathbf{W}, \mathbf{X}) = 1)$.

\section{A Conditional Randomization Test for the Case of Non-Categorical Covariates} \label{s:conditionalRandomizationTest}

The conditional randomization test discussed in Section \ref{ss:conditionalRandomizationTest} is equivalent to a permutation test within $S$ strata. This is analogous to analyzing a completely randomized experiment as if it were a blocked randomized experiment. We follow this intuition by proposing a conditional randomization test that is analogous to analyzing a completely randomized experiment as if it were a rerandomized experiment, where the rerandomization scheme incorporates a general form of covariate balance.

Rerandomization involves randomly allocating units to treatment and control until a certain level of prespecified covariate balance is achieved. Thus, rerandomization requires specifying a metric for covariate balance. We first consider an omnibus measure of covariate balance and the corresponding conditional randomization test. We then extend this conditional randomization test to flexibly incorporate multiple measures of covariate balance, rather than a single omnibus measure, which we find yields more powerful randomization tests.

\subsection{Conditional Randomization Test Using An Omnibus Measure of Covariate Balance} \label{ss:omnibusCriterion}

The most common covariate balance metric used in the rerandomization literature is the Mahalanobis distance \citep{mahalanobis1936generalized}, which is defined as
\begin{align}
	M &\equiv (\overline{\mathbf{X}}_T - \overline{\mathbf{X}}_C)^T \left[ \text{cov}(\overline{\mathbf{X}}_T - \overline{\mathbf{X}}_C) \right]^{-1}(\overline{\mathbf{X}}_T - \overline{\mathbf{X}}_C) \\
	&= \frac{N_T N_C}{N}(\overline{\mathbf{X}}_T - \overline{\mathbf{X}}_C)^T \left[ \text{cov}(\mathbf{X}) \right]^{-1}(\overline{\mathbf{X}}_T - \overline{\mathbf{X}}_C) \label{eqn:md}
\end{align}
where $\overline{\mathbf{X}}_T$ and $\overline{\mathbf{X}}_C$ are $p$-dimensional vectors of the covariate means in the treatment and control groups, respectively, and $\text{cov}(\mathbf{X})$ is the sample covariance matrix of $\mathbf{X}$, which is fixed across randomizations. The derivation for the equality in (\ref{eqn:md}) can be found in \cite{morgan2012rerandomization}. Note that $\overline{\mathbf{X}}_T - \overline{\mathbf{X}}_C = \frac{\mathbf{X}^T \mathbf{W}}{\sum_{i=1}^N W_i} - \frac{\mathbf{X}^T (\mathbf{1} - \mathbf{W})}{\sum_{i=1}^N 1 - W_i}$, and so $M$ is stochastic through $\mathbf{W}$.

We focus on using the Mahalanobis distance for our conditional randomization test because of its widespread use in measuring covariate balance for non-categorical covariates. Note that the Mahalanobis distance is an omnibus measure for balance among the individual covariates as well as their interactions (see, e.g., \cite{stuart2010matching}). Following \cite{hennessy2016conditional}, we define a criterion $\phi(\mathbf{W}, \mathbf{X})$ such that:
\begin{enumerate}
    \item It is asymmetric in treatment and control.\footnote{In particular, we would like the criterion to be able to distinguish between assignments $\mathbf{W}$ where treated units have higher covariate values and $\mathbf{W}$ where control units have higher covariate values. As discussed in \cite{hennessy2016conditional}, this can be useful information to condition on during a randomization test. In contrast, the Mahalanobis distance is symmetric in treatment and control. Asymmetry allows us to condition on the direction of the mean imbalance between treatment and control, in addition to the degree of covariate overlap between the two groups (as measured by the Mahalanobis distance).}
    \item It conditions on the covariate balance being similar to the observed balance for a particular randomization.
\end{enumerate}
To fulfill these two desires, we consider the following criterion for our conditional randomization test:
\begin{align}
    \phi_{b_L, b_U}(\mathbf{W}, \mathbf{X}) = \begin{cases}
        1 &\mbox{ if } b_L \leq M^{obs} \leq b_U \text{ and } \text{sign}(\overline{\mathbf{X}}_{T, j} - \overline{\mathbf{X}}_{C, j}) = \text{sign}(\overline{\mathbf{X}}_{T, j}^{obs} - \overline{\mathbf{X}}_{C, j}^{obs}) \hspace{0.05 in} \forall j =1,\dots,p \\
        0 &\mbox{ otherwise.} \label{eqn:omnibusPhi}
    \end{cases}
\end{align}
The equality of signs for all covariate mean differences addresses the first item above---in particular, it recognizes whether the treatment or control group has higher covariate values---while the bounds $(b_L, b_U)$ address the second item.

The criterion (\ref{eqn:omnibusPhi}) only considers randomizations that correspond to covariate balance similar to the observed $M$. Restricting $M$ to be within the bounds $(b_L, b_U)$ is analogous to stratifying the Mahalanobis distance and restricting $M$ to be in the same stratum as the observed $M$. Now we outline two procedures for selecting $(b_L, b_U)$ for our conditional randomization test.

\subsubsection{How to Choose the Bounds $(b_L, b_U)$} \label{sss:choosingBounds}

To gain some intuition for how to choose the bounds, note that the interval $(b_L, b_U)$ should be narrow enough around the observed $M$ such that the corresponding $\mathbb{W}$ sufficiently conditions on the observed covariate balance, but also the interval should be wide enough such that a powerful randomization test can still be performed. For example, consider the most narrow interval possible, when $b_L = b_U = M^{obs}$. In this case, there may be only a single randomization such that $M = M^{obs}$ (i.e., $|\mathbb{W}| = 1$) and thus our conditional randomization test completely loses its power, even though it is fully conditioning on the observed covariate balance.

We will consider two ways to pick $(b_L, b_U)$, presented as Procedures 1 and 2 below. Procedure 1 selects the bounds unconditionally of $M^{obs}$, while Procedure 2 does the same conditional on $M^{obs}$. In Section \ref{ss:validity} we establish that Procedure 1 yields a valid randomization test, and we also discuss the extent to which Procedure 2 yields a valid randomization test. \\

\noindent
\noindent\fbox{%
\parbox{\textwidth}{%
\textbf{Procedure 1 for Selecting $(b_L,b_U)$: Bin the Mahalanobis Distance}
\begin{enumerate}
	\item Approximate the \textit{sign-constrained} randomization distribution of the Mahalanobis distance by generating randomizations $\mathbf{w}^{(1)},\dots,\mathbf{w}^{(D)}$ such that $\text{sign}(\overline{\mathbf{X}}_{T, j} - \overline{\mathbf{X}}_{C, j}) = \text{sign}(\overline{\mathbf{X}}_{T, j}^{obs} - \overline{\mathbf{X}}_{C, j}^{obs}) \hspace{0.05 in} \forall j =1,\dots,p$, and computing the corresponding $M^{(1)},\dots,M^{(D)}$.
	\item Before observing $M^{obs}$, bin the aforementioned randomization distribution into $C$ categories. Denote the cutoff points for these $C$ bins as $m_1,\dots,m_{C+1}$, where $m_1 \leq \cdots \leq m_{C+1}$ and $m_1 = 0$ and $m_{C+1} = \infty$.
	\item After observing $M^{obs}$, set $b_L = m_c$ and $b_U = m_{c+1}$ for the $c \in \{1,\dots,C\}$ such that $b_L \leq M^{obs} \leq b_U$.
\end{enumerate}
}
} \\ \\

\noindent
\noindent\fbox{%
\parbox{\textwidth}{%
\textbf{Procedure 2 for Selecting $(b_L,b_U)$: Build a Neighborhood around $M^{obs}$}
\begin{enumerate}
	\item Approximate the \textit{sign-constrained} randomization distribution of the Mahalanobis distance by generating randomizations $\mathbf{w}^{(1)},\dots,\mathbf{w}^{(D)}$ such that $\text{sign}(\overline{\mathbf{X}}_{T, j} - \overline{\mathbf{X}}_{C, j}) = \text{sign}(\overline{\mathbf{X}}_{T, j}^{obs} - \overline{\mathbf{X}}_{C, j}^{obs}) \hspace{0.05 in} \forall j =1,\dots,p$, and computing the corresponding $M^{(1)},\dots,M^{(D)}$.
	\item Specify an acceptance probability $p_a \in (0, 1]$ that denotes the proportion of the aforementioned randomization distribution to be included in $(b_L,b_U)$.
	\item After observing $M^{obs}$, let $\mathcal{M}_L$ be the set of $\frac{Dp_a}{2}$ Mahalanobis distances that are immediately below $M^{obs}$, and let $\mathcal{M}_U$ be the set of $\frac{Dp_a}{2}$ Mahalanobis distances that are immediately above $M^{obs}$. Then, set $b_L = \min \mathcal{M}_L$ and $b_U = \max \mathcal{M}_U$.
	\begin{itemize}
		\item If there are fewer than $\frac{Dp_a}{2}$ Mahalanobis distances immediately below $M^{obs}$, set $\mathcal{M}_L$ as the set of all Mahalanobis distances below $M^{obs}$, and set $\mathcal{M}_U$ as the set of Mahalanobis distances immediately above $M^{obs}$ such that $|\mathcal{M}_L| + |\mathcal{M}_U| = D p_a$.
		\item If there are fewer than $\frac{Dp_a}{2}$ Mahalanobis distances immediately above $M^{obs}$, set $\mathcal{M}_U$ as the set of all Mahalanobis distances above $M^{obs}$, and set $\mathcal{M}_L$ as the set of Mahalanobis distances immediately below $M^{obs}$ such that $|\mathcal{M}_L| + |\mathcal{M}_U| = D p_a$. 
	\end{itemize}
\end{enumerate}
}
} \\

Procedure 1 categorizes the Mahalanobis distance and then sets $(b_L,b_U)$ according to the category that $M^{obs}$ falls into. Procedure 2 sets $(b_L, b_U)$ according to the Mahalanobis distances that are immediately around $M^{obs}$, such that $D p_a$ of the Mahalanobis distances $M^{(1)},\dots,M^{(D)}$ are contained in $(b_L, b_U)$, with $M^{obs}$ being the median of $(b_L,b_U)$ (except for the two corner cases noted in the final step of Procedure 2). Furthermore, one can use rejection sampling to generate the randomizations in Step 1 of Procedures 1 and 2: generate a complete randomization $\mathbf{w} \sim P(\mathbf{W})$, where $P(\mathbf{W})$ is defined in (\ref{eqn:completeRandomization}), and only keep $\mathbf{w}$ if the sign constraint is fulfilled by $\mathbf{w}$. In the simulation study discussed in Section \ref{s:simulations}, we focus on Procedure 2, because it ensures that the hypothetical randomizations $\mathbf{w}^{(1)},\dots,\mathbf{w}^{(D)}$ used during the conditional randomization test are the randomizations most similar to the observed one in terms of covariate balance. 

\subsubsection{Rejection-Sampling Approach for Performing the Conditional Randomization Test} \label{sss:rejectionSampling}

The conditional randomization test proceeds according to the three-step procedure in Section \ref{ss:setupProcedureRandomizationTests} after $b_L$ and $b_U$ are specified and the criterion (\ref{eqn:omnibusPhi}) is defined. While we assume that $P(\mathbf{W} | \phi_{b_L, b_U}(\mathbf{W}, \mathbf{X}) = 1)$ is uniformly distributed, random samples from this conditional distribution no longer correspond to random permutations of $\mathbf{W}^{obs}$ as in the unconditional randomization test in Section \ref{ss:unconditionalRandomizationTest} or the conditional randomization test in Section \ref{ss:conditionalRandomizationTest}. Similar to how the randomizations in Step 1 of Procedures 1 and 2 can be generated, we propose a simple rejection-sampling algorithm to generate a random draw from $P(\mathbf{W} | \phi_{b_L, b_U}(\mathbf{W}, \mathbf{X}) = 1)$:
\begin{enumerate}
    \item Generate a random draw $\mathbf{w}$ from $P(\mathbf{W})$ defined in (\ref{eqn:completeRandomization}).
    \item Accept $\mathbf{w}$ if $\phi_{b_L, b_U}(\mathbf{w}, \mathbf{X}) = 1$; otherwise, repeat Step 1.
\end{enumerate}
Note that, as $p_a$ gets smaller, it will be more computationally intensive to generate random samples from $P(\mathbf{W} | \phi_{b_L, b_U}(\mathbf{W}, \mathbf{X}) = 1)$, but it corresponds to more precisely conditioning on the observed covariate balance. If generating random samples from $P(\mathbf{W} | \phi_{b_L, b_U}(\mathbf{W}, \mathbf{X}) = 1)$ via rejection-sampling is computationally intensive, one can use an alternative approach proposed by \cite{branson2018randomization}, which uses importance-sampling to approximate randomization test $p$-values at a lower computational cost than rejection-sampling.

In Section \ref{s:simulations} we show via simulation that this conditional randomization test is more powerful than the standard unconditional randomization test, because the former conditions on a measure of covariate balance. However, the criterion (\ref{eqn:omnibusPhi}) uses an omnibus measure of covariate balance, which may not sufficiently condition on the observed randomization if the number of covariates $p$ is large. We now extend this procedure to more precisely condition on the observed covariate balance for a given randomization by incorporating multiple measures of covariate balance. We show in Section \ref{s:simulations} that this extension results in a further gain in statistical power.

\subsection{Conditional Randomization Test Using Multiple Measures of Covariate Balance} \label{ss:multipleMDsRandomizationTest}

Consider $t = 1, \dots , T$ tiers (or sets) of covariates that are of interest as specified by the researcher. Let $\mathbf{X}^{(t)} \equiv (\mathbf{X}_1^{(t)}, \dots, \mathbf{X}_{k_t}^{(t)})$ 
denote the covariates in tier $t$, where each covariate only appears in one of the $T$ tiers. Then, define
\begin{align}
	M^{(t)} \equiv \frac{N_T N_C}{N} (\bar{\mathbf{X}}_{T}^{(t)} - \bar{\mathbf{X}}_{C}^{(t)})^T [\text{cov}(\mathbf{X}^{(t)})]^{-1} (\bar{\mathbf{X}}_{T}^{(t)} - \bar{\mathbf{X}}_{C}^{(t)})
\end{align}
as the Mahalanobis distance for the covariates in tier $t$. This setup of dividing covariates into tiers is similar to \cite{morgan2015rerandomization}, who developed a rerandomization framework that forces each $M^{(t)}$ to be sufficiently small by design. Note that the setup in Section 3.1 corresponds to $T = 1$ tiers.

Our proposed conditional randomization test follows a procedure similar to that in Section \ref{ss:omnibusCriterion}, but within each tier $t$. Define the criterion
\begin{align}
    \phi^{(t)}(\mathbf{W}, \mathbf{X}) = \begin{cases}
        1 &\mbox{ if } b_{L_t} \leq M^{(t)} \leq b_{U_t} \text{ and } \text{sign}(\overline{\mathbf{X}}_{tj, T} - \overline{\mathbf{X}}_{tj, C}) = \text{sign}(\overline{\mathbf{X}}_{tj, T}^{obs} - \overline{\mathbf{X}}_{tj, C}^{obs}) \hspace{0.05 in} \forall j =1,\dots,k_t \\
        0 &\mbox{ otherwise.} \label{eqn:tierPhi}
    \end{cases}
\end{align}
for some lower and upper bounds $b_{L_t}$ and $b_{U_t}$ for each tier $t$. Then, define the overall criterion
\begin{align}
    \phi_T(\mathbf{W}, \mathbf{X}) = \prod_{t=1}^T \phi^{(t)}(\mathbf{W}, \mathbf{X}) \label{eqn:multipleMDPhi}
\end{align}
The bounds $(b_{L_t}, b_{U_t})$ are chosen separately for each tier using the procedure discussed in Section \ref{sss:choosingBounds}. This requires choosing an acceptance probability $p_{a_t}$ for each tier. Because a smaller $p_{a_t}$ corresponds to more stringent conditional inference, tiers with covariates that are believed to be most relevant to the outcomes should be assigned smaller $p_{a_t}$. However, recall that smaller $p_{a_t}$ corresponds to more computational time required to obtain draws from $P(\mathbf{W} | \phi_T(\mathbf{W}, \mathbf{X}) = 1)$ via our rejection-sampling algorithm discussed in Section \ref{sss:rejectionSampling}.

\subsection{The Validity of Conditional Randomization Tests} \label{ss:validity}

A test is valid if $P(p \leq \alpha | H_0) \leq \alpha$, where $H_0$ is the Sharp Null Hypothesis and $p$ is the calculated $p$-value.
In our context, $p$ is a function of the observed assignment and the testing procedure, and the  probability is taken over the true assignment mechanism with the potential outcomes held fixed.
For our conditional tests, the $p$-value is calculated as the probability of observing a test statistic more extreme than the observed one across randomizations $\mathbf{w}$ such that $\phi_T(\mathbf{w}, \mathbf{X}) = 1$, for a specific $\phi_T(\mathbf{w}, \mathbf{X})$ determined by the observed assignment and covariates.  
Thus, the validity of our conditional randomization test depends on the criterion $\phi_T(\mathbf{W},\mathbf{X})$, which---as shown in (\ref{eqn:tierPhi})---is defined by the bounds in each tier and the covariate sign constraints. 
In Section \ref{sss:choosingBounds}, Procedure 1 defines the bounds before randomization, whereas Procedure 2 defines the bounds based on $\mathbf{W}^{obs}$ after randomization. 
This latter case induces complications to establishing validity that we believe have not been previously discussed in the literature. 
In what follows, we discuss why exact validity may not necessarily hold for the conditional randomization test that uses Procedure 2, and establish validity for the test that uses Procedure 1.

Define $\mathcal{B}$ as the set of possible bounds and $\mathcal{S}$ as the set of possible covariate signs across all randomizations, and define $\mathbb{W}_{b,s}$ as the set of all randomizations that lead to particular bounds $\mathbf{b} \in \mathcal{B}$ and signs $\mathbf{s} \in \mathcal{S}$. 
The collection of $\mathbb{W}_{b,s}$ partition $\mathbb{W}$ into non-overlapping sets.\footnote{Recall that $\mathbb{W}$ is the set of treatment assignments with positive probability; see section \ref{s:reviewRandomizationTests}.} 
The overall probability of our conditional randomization test falsely rejecting the null can then be decomposed as
\begin{align}
	P(p \leq \alpha | H_0) &= \sum_{b \in \mathcal{B}} \sum_{s \in \mathcal{S}} P(p \leq \alpha | H_0, \mathbf{W} \in \mathbb{W}_{b,s}) P( \mathbf{W} \in \mathbb{W}_{b,s} ),
\end{align}
Given the above, a sufficient condition for establishing validity is that $P(p \leq \alpha | H_0, \mathbf{W} \in \mathbb{W}_{b,s}) \leq \alpha$ for all $\mathbf{b} \in \mathcal{B}$ and $\mathbf{s} \in \mathcal{S}$.

A given $b$ and $s$ pair specify a specific conditioning function $\phi_{T}(\mathbf{W},\mathbf{X})$.
Let $\mathbb{W}_\phi \equiv \{\mathbf{W}: \phi_{T}(\mathbf{W},\mathbf{X}) = 1\}$ be the set of randomizations satisfying a given function $\phi_{T}(\cdot)$.
Then, our calculated $p$-value, conditioned on an observed randomization, consequent $\phi_{T}(\cdot)$, and outcome will be
\begin{align}
	p \equiv \sum_{\mathbf{w} \in \mathbb{W}_{\phi}} \mathbb{I}(|t(Y(\mathbf{w}), \mathbf{w}, \mathbf{X})| \geq |t^{obs}|) P(\mathbf{W} = \mathbf{w} | \mathbf{W} \in \mathbb{W}_\phi ) ,\label{eqn:phiPValue} 
\end{align}
where $t^{obs} \equiv t(\mathbf{y}^{obs}, \mathbf{W}^{obs}, \mathbf{X})$.

Under the null, $\mathbf{y}^{obs}$ and $\mathbf{X}$ are both invariant to random assignment, making our test statistic solely a function of $\mathbf{W}$.
Under the null, then, let $U_{\phi}$ be a random variable whose distribution is that of $|t(\mathbf{y}^{obs}, \mathbf{w}, \mathbf{X})|$, where $\mathbf{w}$ is uniformly distributed across the elements of $\mathbb{W}_{\phi}$, and let $U_{b,s}$ be analogously defined for $\mathbb{W}_{b,s}$. 
(Note that $\mathbf{W}^{obs} \in \mathbb{W}_{b,s}$, because the realized $\mathbf{b}$ and $\mathbf{s}$ are specified by $\mathbf{W}^{obs}$.)
Now consider our conditional probability $P(p \leq \alpha | H_0, \mathbf{W} \in \mathbb{W}_{b,s})$ for some specific $b$ and $s$.
Given this conditioning, our original test statistic is distributed as $U_{b,s}$.
Regardless of the observed value of our test statistic, we have that our reference distribution will be $U_\phi$, for our given $\phi_{T}(\cdot)$.
Thus, our $p$-value, conditioned on our original randomization giving us our given $b$, $s$ pair will then be the upper tail of our reference distribution, calculated as
\begin{align}
	1 - F_{U_{\phi}}(U_{b,s}), \label{eqn:cdf}
\end{align}
where $F_{U_{\phi}}(\cdot)$ is the cumulative distribution function of $U_{\phi}$. 
Here $U_\phi$ is a function of $b$ and $s$, given the potential outcomes and covariates.

Typically, validity of a randomization test is proven by arguing that $p$-values of the form (\ref{eqn:cdf}) are uniformly distributed by applying the probability integral transform (for an example of this method of proof, see \citealt[Section 2]{hennessy2016conditional}). 
When Procedure 1 is used to select the bounds, $\mathbb{W}_{\phi} = \mathbb{W}_{b,s}$; i.e., all of the assignments used in the conditional randomization test are the same assignments that would lead to the realized $\mathbf{b}$ and $\mathbf{s}$. 
Therefore, $U_{b,s}$ and $U_{\phi}$ have the same distribution under Procedure 1, and validity immediately follows from (\ref{eqn:cdf}). 
However, because Procedure 2 specifies the bounds as a neighborhood around $M^{obs}$, the conditional randomization test under Procedure 2 uses randomizations that may not have led to the realized $\mathbf{b}$ and $\mathbf{s}$. 
As a result, $\mathbb{W}_{\phi}$ and $\mathbb{W}_{b,s}$ will differ, and $U_{b,s}$ and $U_{\phi}$ will not necessarily have the same distribution. 
Consequently, our conditional randomization test that uses Procedure 2 for selecting the bounds is not necessarily valid. 
Nonetheless, in Section \ref{s:simulations} and the Appendix, we find that our conditional randomization test using Procedure 2 is empirically valid under a wide variety of scenarios.
This in part stems from the centering of our reference distributions around the test statistics; by contrast, if we had always selected distributions less extreme than the observed, we could induce invalidity.
We leave investigating when validity formally holds when randomization test $p$-values are of the form (\ref{eqn:cdf}) for two differing distributions as a promising line for future research.

\section{Simulation Study: Conditional and Unconditional Performance of Conditional and Unconditional Randomization Tests} \label{s:simulations}

We now conduct a simulation study to explore the statistical power of the unconditional randomization test from Section \ref{ss:unconditionalRandomizationTest}, our conditional randomization tests from Sections \ref{ss:omnibusCriterion} and \ref{ss:multipleMDsRandomizationTest}, and another conditional randomization test inspired by Coarsened Exact Matching (CEM). CEM was designed for observational studies to find a subset of treatment and control units that match exactly on a coarsened covariate space \citep{iacus2011multivariate,iacus2012causal}. Even though CEM was developed for observational studies and not randomization tests, we include it in our comparison because---as we noted at the end of Section \ref{s:reviewRandomizationTests}---coarsening $\mathbf{X}$ into strata is one option for performing a conditional randomization test in the face of continuous covariates. Thus, it is the most natural test to compare to our conditional randomization test.

In what follows, we find that our conditional randomization test using $\hat{\tau}_{sd}$ is more powerful than the unconditional randomization test using $\hat{\tau}_{sd}$ as well as the CEM-based tests. Furthermore, we find that our test is comparable to an unconditional randomization test using a regression-based test statistic. Finally, we find that the conditional randomization tests and an unconditional randomization test using a regression-based test statistic are all valid both unconditionally and conditional on the data, whereas the unconditional randomization test that uses an unadjusted test statistic is only valid unconditionally.

\subsection{Simulation Procedure} \label{ss:simulationProcedure}

Consider $N = 100$ units whose potential outcomes are generated according to the following model:
	\begin{equation}
	\begin{aligned}
		&Y_i(0) | X_i = \beta(0.1X_{i1} + 0.2X_{i2} + 0.3X_{i3} + 0.4X_{i4}) + \epsilon_i, \hspace{0.1 in} i=1,\dots,100 \\
		&Y_i(1) = Y_i(0) + \tau \label{eqn:potentialOutcomesModel}
	\end{aligned}
	\end{equation}
    where $X_{i1}, X_{i2},X_{i3}, X_{i4},$ and $\epsilon_i$ are independently and randomly sampled from a $N(0,1)$ distribution. The parameters $\beta$ and $\tau$ take on values $\beta \in \{0, 1.5, 3\}$ and $\tau \in \{0, 0.1, \dots 1\}$ across simulations. As $\beta$ increases, the covariates become more associated with the outcome; as $\tau$ increases, the treatment effect increases and thus should be easier to detect.
	
    Once the potential outcomes are generated, units are randomized to treatment and control such that $N_T = 50$ units receive treatment and $N_C = 50$ units receive control; in other words, units are assigned according to the completely randomized assignment mechanism (\ref{eqn:completeRandomization}). This is repeated such that 1,000 randomizations are produced using the same fixed potential outcomes. In the Appendix we also consider an unbalanced design where an unequal number of units are assigned to treatment and control; however, the results for that scenario are largely the same as the results presented here, where $N_T = N_C = 50$.

    For each randomization, five separate randomization tests were performed:
	\begin{enumerate}
		\item \textbf{Unconditional Randomization Test}: The procedure described in Section \ref{ss:unconditionalRandomizationTest}, using the test statistic $\hat{\tau}_{sd}$ given in (\ref{eqn:meanDiffEst}).
        \item \textbf{Conditional Randomization Test}: The procedure described in Section \ref{ss:multipleMDsRandomizationTest} using the criterion (\ref{eqn:multipleMDPhi}), which requires specifying the number of covariate tiers $T$ and acceptance probability $p_a$ in Procedure 2 for selecting the bounds within each tier. We consider number of tiers $T \in \{1, 2, 4\}$ and acceptance probabilities $p_{a} \in \{0.1, 0.25, 0.5\}$. The $T = 1$ case corresponds to the procedure described in Section \ref{ss:omnibusCriterion}.\footnote{For $T = 2$, the first two covariates are in one tier while the last two are in another tier. For $T = 4$, all covariates are in their own tier.} For each tier, we choose $(b_{L_t}, b_{U_t})$ by setting all tier-level acceptance probabilities $p_{a_t}$ to be equal, where the overall acceptance probability is $p_a = \prod_{t=1}^T p_{a_t}$.\footnote{Note that this equality holds only because the covariates in each tier are independent. Thus, $p_{a_t} = (p_a)^{1/T}$ for all tiers $t = 1,\dots,T$.} We use the test statistic $\hat{\tau}_{sd}$.
		\item \textbf{Unconditional Randomization (with model-adjusted test statistic)}: The procedure described in Section \ref{ss:unconditionalRandomizationTest}, using the test statistic $\hat{\tau}_{int}$, which is defined as the estimated coefficient for $W_i$ from the linear regression of $Y_i$ on $W_i$, $\mathbf{x}_i$, and $W_i(\mathbf{x}_i - \overline{\mathbf{X}})$. This test statistic was discussed in \cite{lin2013agnostic}, but within the context of Neymanian inference rather than randomization tests.
		\item \textbf{Coarsened Exact Matching (Prespecified Groups)}: Each covariate is coarsened into $G$ groups according to the quantiles of $N(0,1)$, thus coarsening the $\mathbb{R}^4$ covariate space into $G^4$ strata. Then, to perform a randomization test using $\hat{\tau}_{sd}$ as a test statistic, $\mathbf{W}^{obs}$ is permuted many times within each stratum. We consider number of groups $G \in \{2,3,4\}$.
		\item \textbf{Coarsened Exact Matching (Automatic Groups)}: The same as the previous test, but the $G$ groups are chosen automatically by the \texttt{R} function \texttt{cem}.
	\end{enumerate}
Our motivation for including the third randomization test in our comparison is that \cite{hennessy2016conditional} found that their conditional randomization test using $\hat{\tau}_{sd}$ is comparable to the unconditional randomization test using $\hat{\tau}_{ps}$ defined in (\ref{eqn:postStratEst}), and that $\hat{\tau}_{ps}$ is equivalent to $\hat{\tau}_{int}$ when covariates are categorical \citep{lin2013agnostic}. We also considered our conditional randomization test using $\hat{\tau}_{int}$ instead of $\hat{\tau}_{sd}$, and found that the power results for that test are essentially the same as those for the unconditional randomization test using $\hat{\tau}_{int}$; we relegate those results to the Appendix.

Meanwhile, the last two procedures utilize CEM. These conditional tests are identical to the test of \cite{hennessy2016conditional} using the assignment mechanism (\ref{eqn:strataRandomization}), where the strata are chosen via CEM. In the CEM (Prespecified Groups) procedure, the strata are specified according to the quantiles of the known distributions of the covariates. Meanwhile, in the CEM (Automatic Groups) procedure, the strata are automatically specified according to Sturges' rule, which uses the range of the covariates and is the default option in the \texttt{cem} \texttt{R} package \citep{iacus2009cem}. Details about this procedure and other automated procedures in the context of CEM are discussed in \cite{iacus2012causal}.

\subsection{Simulation Results: Unconditional Performance} \label{ss:unconditionalProperties}

\begin{figure}
	\centering
	\begin{subfigure}{\textwidth}
	\centering
 	\includegraphics[scale=0.5]{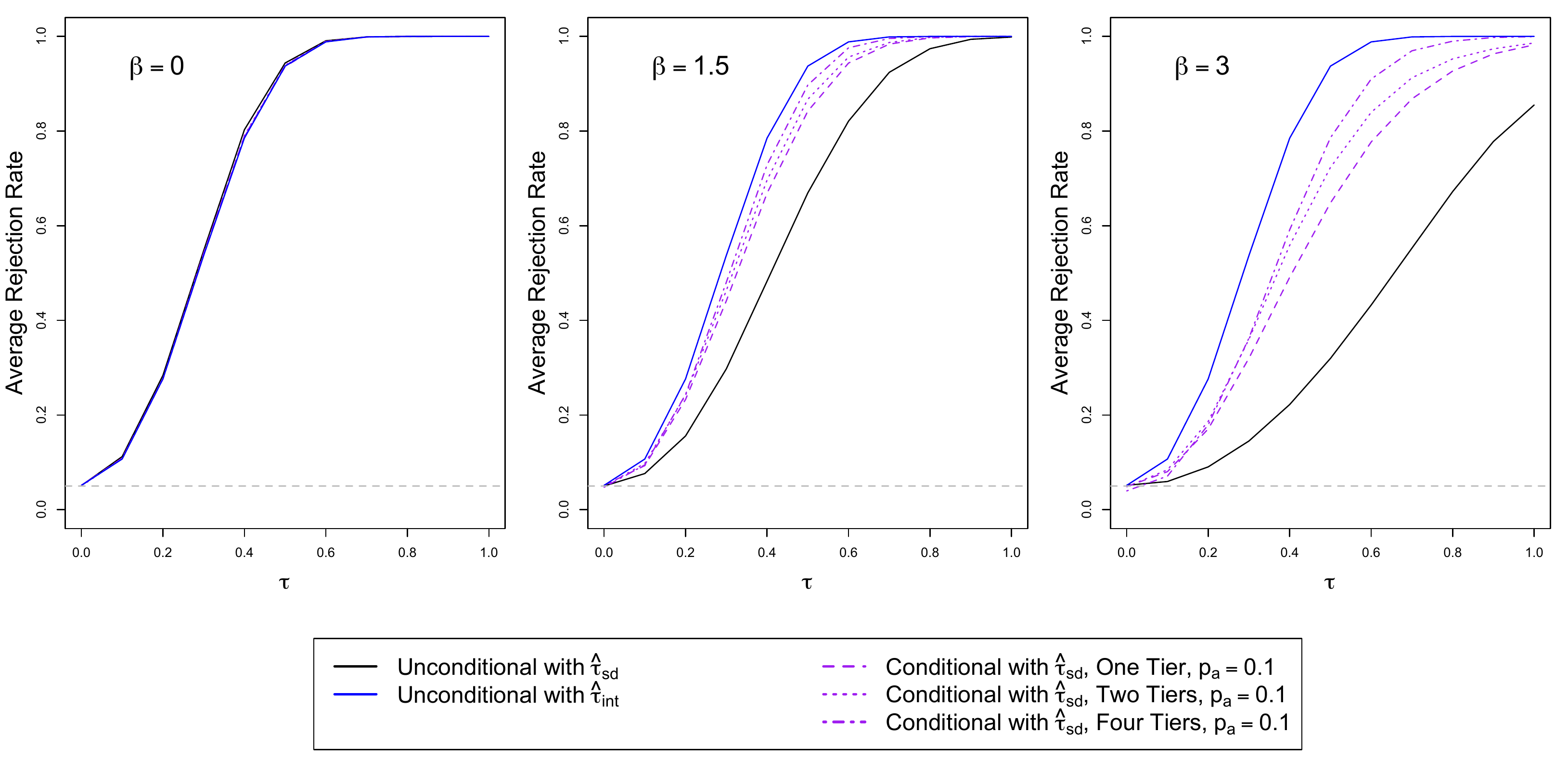}
 	\caption{For the conditional randomization test, different tiers and a fixed $p_a = 0.1$ acceptance probability.}
 	\label{fig:powerPlotDifferentTiers}
 	\end{subfigure}
 	\begin{subfigure}{\textwidth}
    \centering
    \includegraphics[scale=0.5]{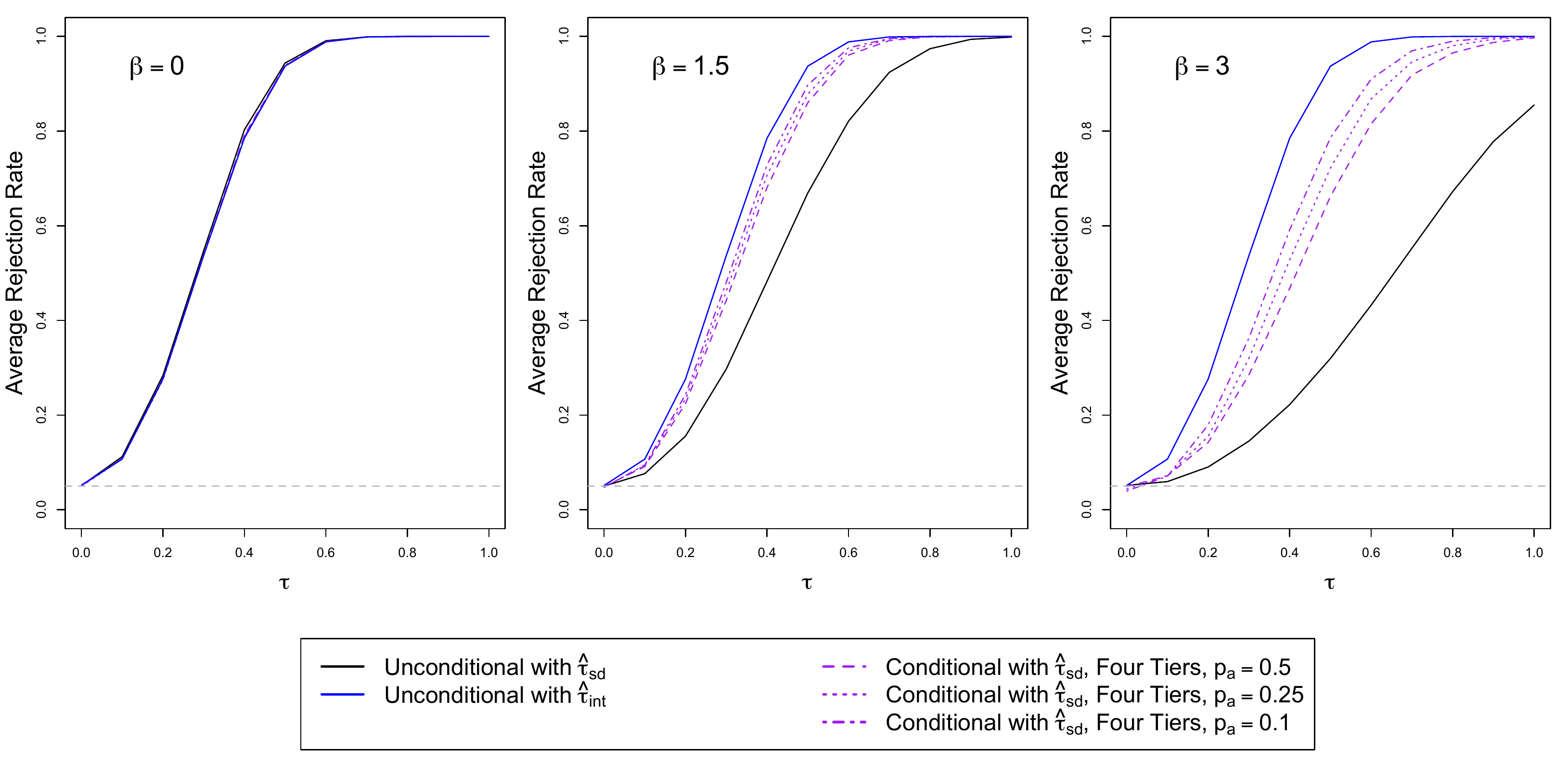}
    \caption{For the conditional randomization test, different acceptance probabilites and a fixed $T = 4$ tiers.}
    \label{fig:powerPlotDifferentPa}
 \end{subfigure} 
 \caption{Average rejection rate (power) of Fisher's Sharp Null for the unconditional randomization test using $\hat{\tau}_{sd}$ and $\hat{\tau}_{int}$, as well as our conditional randomization test using $\hat{\tau}_{sd}$.}
 \label{fig:powerPlot}
 \end{figure}

We first assess statistical power, which corresponds to how often each randomization test rejected Fisher's Sharp Null across the 1,000 complete randomizations when $\tau > 0$. The average rejection rates for the unconditional randomization tests using $\hat{\tau}_{sd}$ and $\hat{\tau}_{int}$ as well as our conditional randomization test are presented in Figure \ref{fig:powerPlot} for various values of $\beta$ and $\tau$. Figure \ref{fig:powerPlotDifferentTiers} displays results for a fixed acceptance probability $p_a = 0.1$ and different numbers of tiers, while Figure \ref{fig:powerPlotDifferentPa} displays results for a fixed number of $T = 4$ tiers and different acceptance probabilities.

Several conclusions can be made from Figure \ref{fig:powerPlot}. First, when $\beta = 0$ (i.e., when the covariates are not associated with the outcome), all of the randomization tests are essentially equivalent. When the covariates are associated with the outcome, our conditional randomization test is more powerful than the unconditional randomization test that uses $\hat{\tau}_{sd}$. Furthermore, the power of our conditional randomization test increases as the acceptance probability $p_a$ decreases and/or the number of tiers increases; this is expected: lower $p_a$ and higher $T$ corresponds to more stringent conditioning. 

Figure \ref{fig:powerPlotDifferentTiers} suggests that practitioners can increase power by increasing the number of tiers without any additional computational cost (i.e., without decreasing the acceptance probability). Furthermore, Figure \ref{fig:powerPlotDifferentPa} suggests that the \textit{additional} gain in power decreases as $p_a$ decreases, which echoes the observation made by \cite{li2018asymptotic} in the rerandomization literature that the marginal benefit to decreasing $p_a$ decreases as $p_a$ decreases. Analogous figures for the $T = 1$ and $T = 2$ cases are in the Appendix; by comparing those figures with Figure \ref{fig:powerPlotDifferentPa}, it can be seen that the additional gain in power from decreasing $p_a$ increases as $T$ increases. This observation emphasizes the benefits of conditioning on multiple measures of covariate balance rather than a single omnibus measure. Further discussion on this point is in the Appendix.

Meanwhile, Figure \ref{fig:powerPlot} also shows that the unconditional randomization test using $\hat{\tau}_{int}$ was more powerful than all of the conditional and unconditional randomization tests using $\hat{\tau}_{sd}$. However, as $p_a$ gets smaller and $T$ gets larger---i.e., as conditioning becomes more stringent---the performance of our conditional randomization test appears to approach that of the unconditional randomization test that uses $\hat{\tau}_{int}$. This reinforces the claim made by \cite{li2018asymptotic} that---in a Neymanian inference context---$\hat{\tau}_{int}$ under complete randomization is equivalent to $\hat{\tau}_{sd}$ under very stringent rerandomization. However, \cite{li2018asymptotic} made this claim about the rerandomization scheme that uses an omnibus measure of covariate balance; our findings suggest that this claim should be qualified to state that the equivalence between $\hat{\tau}_{int}$ under complete randomization and $\hat{\tau}_{sd}$ under rerandomization holds when the rerandomization scheme incorporates separate measures of balance for each covariate used in $\hat{\tau}_{int}$, rather than a single omnibus measure.

Here, $\hat{\tau}_{int}$ is correctly specified because the potential outcomes are generated from a linear model, and one may wonder how the unconditional randomization test using $\hat{\tau}_{int}$ performs when this model is misspecified. We consider this in the Appendix and obtain findings very similar to those presented here. In particular, for the simulation settings considered, we find that it is still beneficial to use the unconditional randomization test with $\hat{\tau}_{int}$ or our conditional randomization test with $\hat{\tau}_{sd}$ in the misspecified case as long as the functions of the covariates used in the regression to construct $\hat{\tau}_{int}$ are correlated with the response; when they are not correlated, these tests are essentially equivalent. In the Appendix we also explore a variety of additional simulation scenarios---when the covariates have positive and negative effects on the potential outcomes, when there are heterogeneous treatment effects, and when the covariates are not normally distributed---and we again find results that are very similar to the results presented here. This suggests that these results hold under a wide variety of scenarios.

\begin{figure}[H]
    \centering
    \begin{subfigure}{\textwidth}
    \centering
    \includegraphics[scale=0.5]{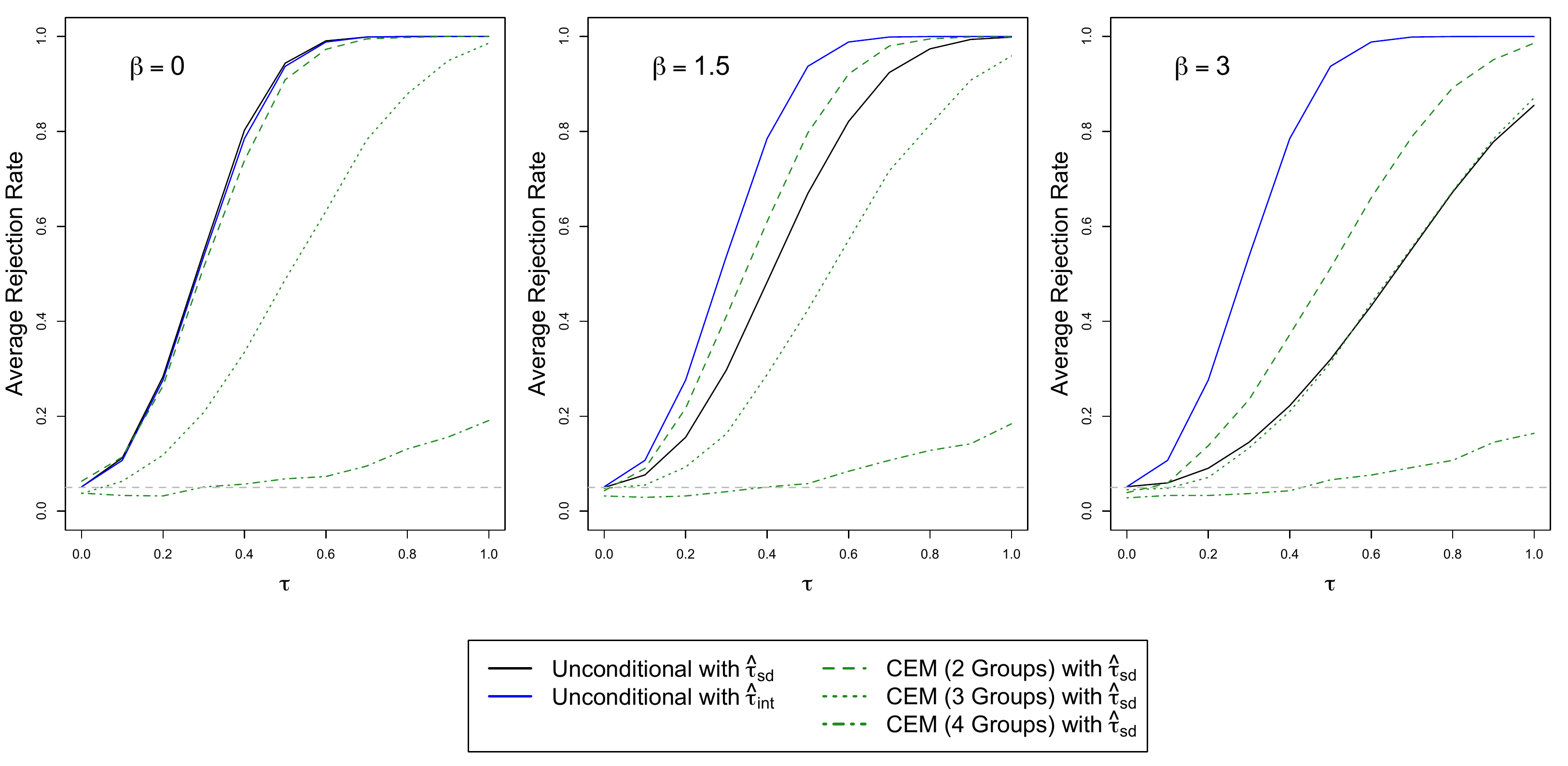}
    \caption{CEM (Prespecified Groups).}
    \label{fig:cemPowerPlot}
 \end{subfigure} 
  \begin{subfigure}{\textwidth}
    \centering
    \includegraphics[scale=0.5]{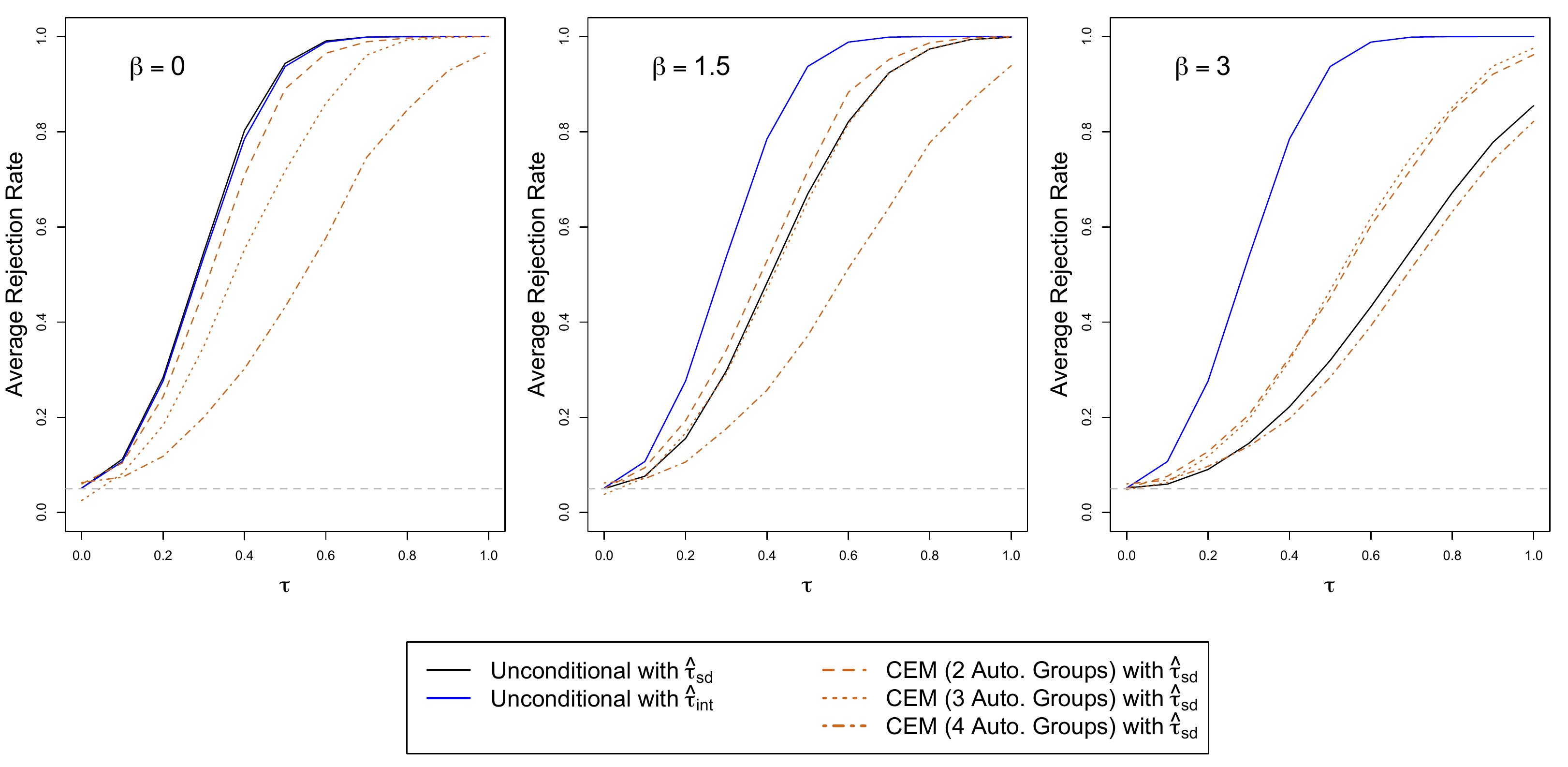}
    \caption{CEM (Automatic Groups).}
    \label{fig:cemAutoPowerPlot}
    \end{subfigure}
    \caption{Average rejection rate (power) of Fisher's Sharp Null for the CEM (Prespecified Groups) and CEM (Automatic Groups) procedures for various number of groups for each covariate. Also shown are results for the unconditional tests using $\hat{\tau}_{sd}$ and $\hat{\tau}_{int}$, which are the same results from Figure \ref{fig:powerPlot}.}
    \label{fig:cemPowerPlots}
 \end{figure}

Now we assess the performance of the conditional randomization tests that use CEM. Figure \ref{fig:cemPowerPlots} shows the average rejection rate of Fisher's Sharp Null for the CEM-based randomization tests. To anchor our comparison, Figure \ref{fig:cemPowerPlots} also includes the results for the unconditional randomization tests using $\hat{\tau}_{sd}$ and $\hat{\tau}_{int}$ (i.e., the same results presented in Figure \ref{fig:powerPlot}). When the covariate space for each covariate is coarsened into $G = 2$ groups, these conditional randomization tests are more powerful than the unconditional randomization test using $\hat{\tau}_{sd}$ when $\beta > 0$, although they are not as powerful as our conditional randomization test or the unconditional randomization test using $\hat{\tau}_{int}$. When the number of groups for each covariate is increased, the power of the conditional randomization tests tend to decrease, especially for the CEM procedure that specifies strata according to the quantiles of the known covariate distributions. At first this finding may be surprising, because more groups should correspond to more stringent conditioning and thus possibly higher power. However, as the number of groups increases, there are fewer strata with both treatment and control units, and thus more units are discarded and there are fewer possible randomizations used during the randomization test. For example, for the CEM (Prespecified Groups) procedure, when there were $G = 2$ groups, on average 4 of the 100 units were discarded across the 1,000 randomizations; when $G = 3$, on average 54 of the 100 units were discarded; and when $G = 4$, on average 88 of the 100 units were discarded. In the most extreme case, if we let the number of groups go to infinity---i.e., not coarsen the continuous covariate space at all---there would not be any treatment and control units with the same covariate values, and thus all units would be discarded.

Meanwhile, there is not a clear winner between the CEM (Prespecified Groups) and CEM (Automatic Groups) procedures, although the CEM (Automatic Groups) procedure is not as severely underpowered for the $G = 4$ case as the CEM (Prespecified Groups) procedure. In their development of CEM, \cite{iacus2009cem,iacus2011multivariate,iacus2012causal} recommend researchers use context-specific knowledge for specifying strata rather than automated procedures, but it is unclear if this should be the recommendation when using CEM for conditional randomization tests. Indeed, the CEM (Prespecified Groups) procedure uses the most context-specific knowledge possible (the actual data-generating process for the covariates), but it does not necessarily perform as well as the automated procedure.

In summary, these findings suggest that it is beneficial to condition on forms of covariate balance that account for continuous covariates, rather than condition on a coarsened version of the continuous covariate space. Furthermore, our methodology allows researchers to condition on the data at hand in a way that increases the power of randomization tests, while coarsening the covariate space may lead to a lack of possible treatment assignments to perform a powerful randomization test.

\subsection{Simulation Results: Conditional Performance} \label{ss:conditionalProperties}

We next examine the performance of the five tests across randomizations that are particularly balanced or imbalanced. First, we generated the potential outcomes using model (\ref{eqn:potentialOutcomesModel}) with $\tau = 0$ (which corresponds to no treatment effect) and $\beta = 3$ (which corresponds to a strong association between the covariates and potential outcomes). Then, we generated 10,000 randomizations and divided these randomizations into 10 groups according to quantiles of the Mahalanobis distance. Thus, the first group consists of the 1,000 best randomizations according to the Mahalanobis distance, while the tenth group consists of the 1,000 worst randomizations. Now we consider whether the five randomization tests are valid (i.e., reject Fisher's Sharp Null when it is true 5\% of the time) for randomizations conditional on a particular level of covariate balance. Conditional validity assesses to what extent these tests are valid across randomizations that are similar to the observed randomization. 

\begin{figure}
		\centering
	\includegraphics[scale=0.65]{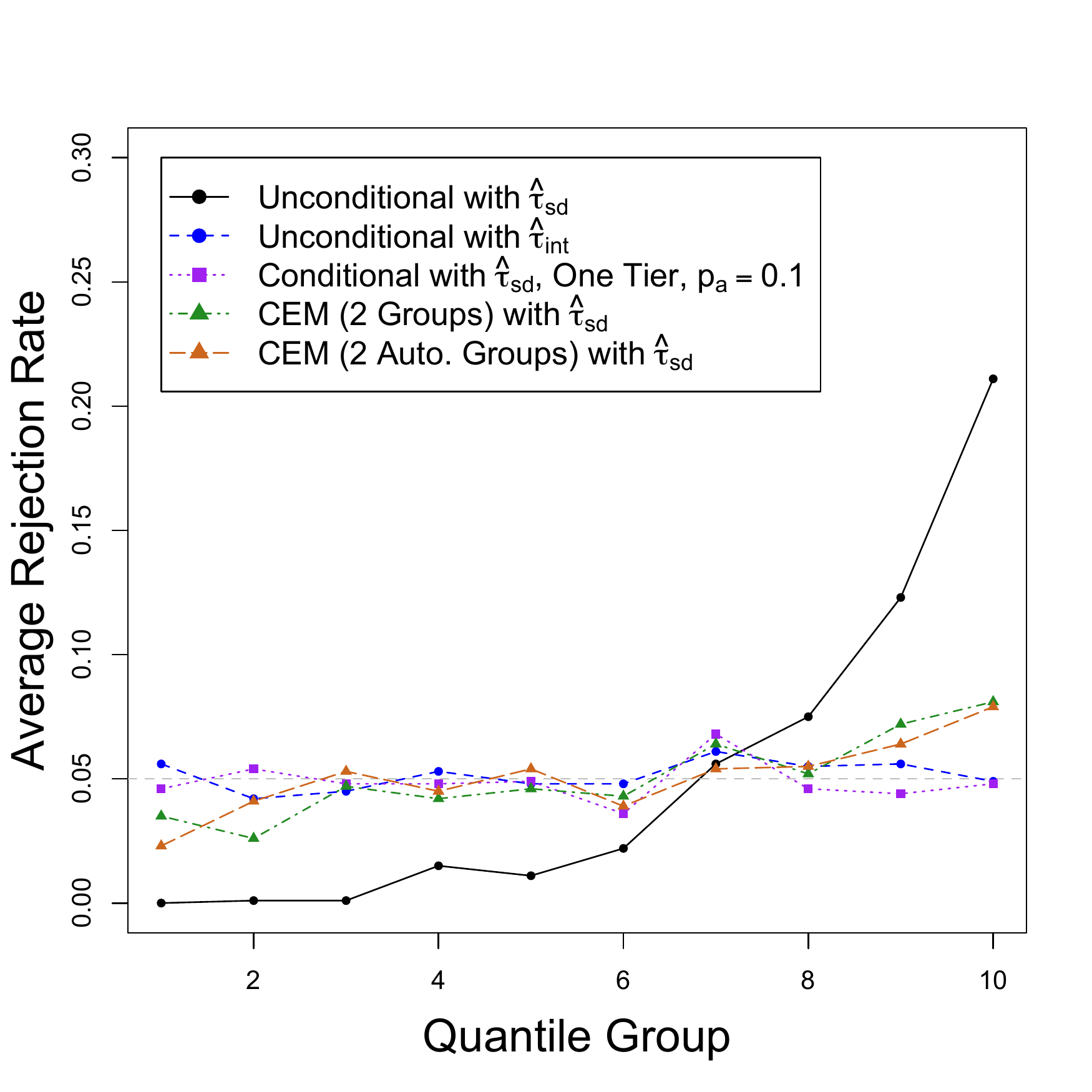}
	\caption{The rejection rate of the five randomization tests when Fisher's Sharp Null Hypothesis is true (i.e., $\tau = 0$) and $\beta = 3$. Rejection rates are shown within each quantile group of the Mahalanobis distance, such that each quantile group corresponds to 1,000 randomizations.}
	\label{fig:conditionalProperties}
\end{figure}

Figure \ref{fig:conditionalProperties} displays the average rejection rate of each randomization test for each of the 10 quantile groups of the Mahalanobis distance. For the CEM-based tests, we display results for $G = 2$ groups, because this resulted in the most power in Section \ref{ss:unconditionalProperties}. The conditional performance for higher groups are similar. Our conditional randomization test that uses $\hat{\tau}_{sd}$ and the unconditional randomization test that uses $\hat{\tau}_{int}$ both exhibit average rejection rates close to the 5\% level across all quantile groups, which suggests that both tests are conditionally valid across randomizations of any particular balance level. The story is quite different for the unconditional randomization test that uses $\hat{\tau}_{sd}$: for low levels of covariate imbalance, the average rejection rate is below the 5\% level, while for high levels of covariate imbalance the average rejection rate is notably above the 5\% level. These rejection rates average out to 5\%---as can be seen in Figure \ref{fig:powerPlot}---and thus the unconditional randomization test that uses $\hat{\tau}_{sd}$ is unconditionally valid, but---as can be seen in Figure \ref{fig:conditionalProperties}---it is not conditionally valid conditional on a particular balance level. In particular, the false rejection rate for the unconditional randomization test that uses $\hat{\tau}_{sd}$ appears to be monotonically increasing in covariate imbalance, which is intuitive given that treatment effects will be increasingly confounded with covariate effects as covariate imbalance increases. Meanwhile, the false rejection rate for the CEM-based tests also appears to be monotonically increasing in covariate imbalance according to the Mahalanobis distance, but to a much less severe degree. This is likely because these tests condition on balance for a coarsened version of the covariate space instead of balance for the continuous covariate space as measured by the Mahalanobis distance. In short, they are conditionally valid for the coarsened covariate space but not the continuous covariate space.

In summary, statistically powerful randomization tests can be constructed by conditioning on covariate balance through the assignment mechanism or by using a model-adjusted test statistic; either option will result in a more powerful test than an unconditional randomization test that uses an unadjusted test statstic. We also find that our conditional randomization test using unadjusted test statistics or unconditional randomization tests using model-adjusted test statistics appear to be approximately equivalent, both across complete randomizations as well as across randomizations of a particular balance level. Furthermore, we find that our conditional randomization test that directly conditions on group-level balance for continuous covariates is more powerful than other conditional randomization tests that condition on a coarsened version of the covariate space. Finally, it is particularly important to condition on group-level covariate balance or use a model-adjusted test statistic to ensure validity across randomizations of a particular balance level, because covariate imbalances can break the conditional validity of unconditional randomization tests that use unadjusted test statistics.

\section{Discussion and Conclusion} \label{s:discussionAndConclusion}

\cite{hennessy2016conditional} outlined a conditional randomization test that conditions on the covariate balance observed after an experiment has been conducted, and showed that these tests are more powerful than standard unconditional randomization tests and comparable to randomization tests that use model-adjusted estimators, such as the post-stratified estimator in \cite{miratrix2013adjusting}. However, \cite{hennessy2016conditional} focused on the case when there are only categorical covariates. Here we proposed a methodology for conducting a randomization test that conditions on a form of covariate balance that allows for non-categorical covariates.

Through simulation, we found that our conditional randomization test is more powerful than unconditional randomization tests that use unadjusted test statistics as well as other conditional randomization tests inspired by the observational study literature, and that it is approximately equivalent to an unconditional randomization test that uses a regression-based test statistic. 
We also found that the conditional randomization tests and the unconditional randomization tests that use adjusted test statistics appear valid conditional on the observed covariate balance; the more traditional unconditional randomization tests that use unadjusted test statistics, however, are clearly not.

The above findings hold under a variety of data-generating scenarios, such as ones with treatment effect heterogeneity or model misspecification. Most of the literature has focused on increasing the power of randomization tests through the choice of the test statistic; to our knowledge, we are the first to do the same through the choice of the assignment mechanism for the general case when non-categorical covariates are present. 
Furthermore, we found evidence that these two avenues for constructing randomization tests are approximately equivalent in terms of statistical power. 
Thus, our methodology can achieve the power of model-adjustment while preserving the transparency of an unadjusted treatment effect estimate, thereby taking advantage of the benefits of both adjusted and unadjusted estimators as discussed by \cite{lin2013agnostic}. 
Relatedly, we also discussed how this finding suggests connections between regression-based estimators after complete randomization and unadjusted estimators after rerandomization, which refines observations previously made by \cite{li2018asymptotic}. 

We focused on randomization tests for randomized experiments, but we believe that this work has implications beyond tests and experiments. Randomization tests can be inverted to yield confidence intervals for treatment effects \citep{rosenbaum2002observational,imbens2015causal}, and thus our method can go beyond testing the presence of a treatment effect. Some have criticized such randomization-based confidence intervals because they commonly make the assumption of a constant treatment effect for all units. However, recent works have suggested how to incorporate treatment effect heterogeneity in randomization tests (e.g., \citealt{ding2016randomization,caughey2016beyond}), and our work adds to this literature by suggesting how forms of covariate balance can be incorporated in randomization tests as well. An interesting line of future work would be to combine our conditional randomization test with these works to conduct randomization-based inference that incorporates both treatment effect heterogeneity and covariate balance.

Furthermore, most work on randomization tests for observational studies has focused on cases where only categorical covariates are present \citep{rosenbaum1984conditional,rosenbaum1988permutation,rosenbaum2002covariance,rosenbaum2002observational}. Our work suggests a way to conduct randomization-based inference for observational studies when non-categorical covariates are present. However, because the assignment mechanism in an observational study is unknown, researchers need to determine when certain assignment mechanisms can be assumed within an observational study before conducting randomization-based inference. See \cite{branson2018my} for a framework for how to conduct conditional randomization-based inference in this context.

\newpage

\section{Appendix: Additional Simulation Results}

Here we present further power results of randomization tests similar to those presented in Section \ref{s:simulations}. All of the following sections and figures discuss the average rejection rate of Fisher's Sharp Null for various randomization tests. In Section \ref{ss:oneTwoTierIntSims}, we consider the same setup discussed in Section \ref{s:simulations} and present results for our conditional randomization test for various acceptance probabilities and one or two tiers (instead of four tiers), as well as results for our conditional randomization test using the regression-adjusted test statistic $\hat{\tau}_{int}$ (instead of $\hat{\tau}_{sd}$). Then, in Sections \ref{ss:alternativeLinearModels} and \ref{ss:misspecifiedModels} we consider other data-generating processes not explored in Section \ref{s:simulations}, including:
\begin{enumerate}
	\item when some covariate effects are positive and some are negative,
	\item when there is treatmenet effect heterogeneity,
	\item when there are non-normal covariates,
	\item when the linear regression used in $\hat{\tau}_{int}$ is misspecified.
\end{enumerate}
The results for the first three are quite similar to the results presented in Section \ref{s:simulations}, and so we discuss them together in Section \ref{ss:alternativeLinearModels}. We discuss results for the misspecified case in Section \ref{ss:misspecifiedModels}.

\subsection{Simulation Results for One and Two Tiers and for Conditional Randomization using $\hat{\tau}_{int}$} \label{ss:oneTwoTierIntSims}

\begin{figure}[H]
    \centering
    \begin{subfigure}[t]{\textwidth}
    \centering
    \includegraphics[scale=0.4]{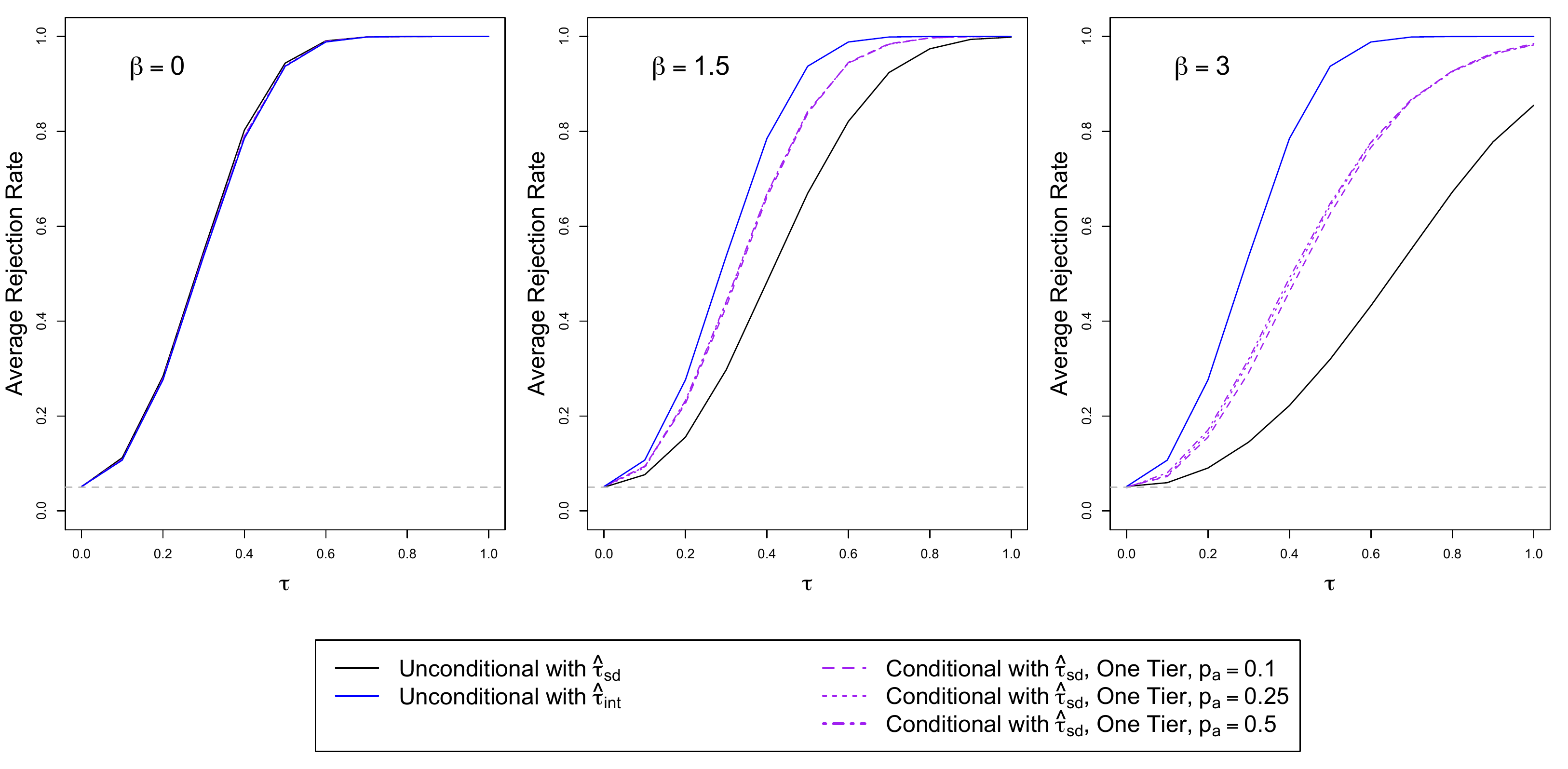}
    \caption{Power results of our conditional randomization test using one tier.}
    \label{fig:powerPlotDifferentPaOneTier}
    \end{subfigure} 

  \begin{subfigure}[t]{\textwidth}
  \centering
    \includegraphics[scale=0.4]{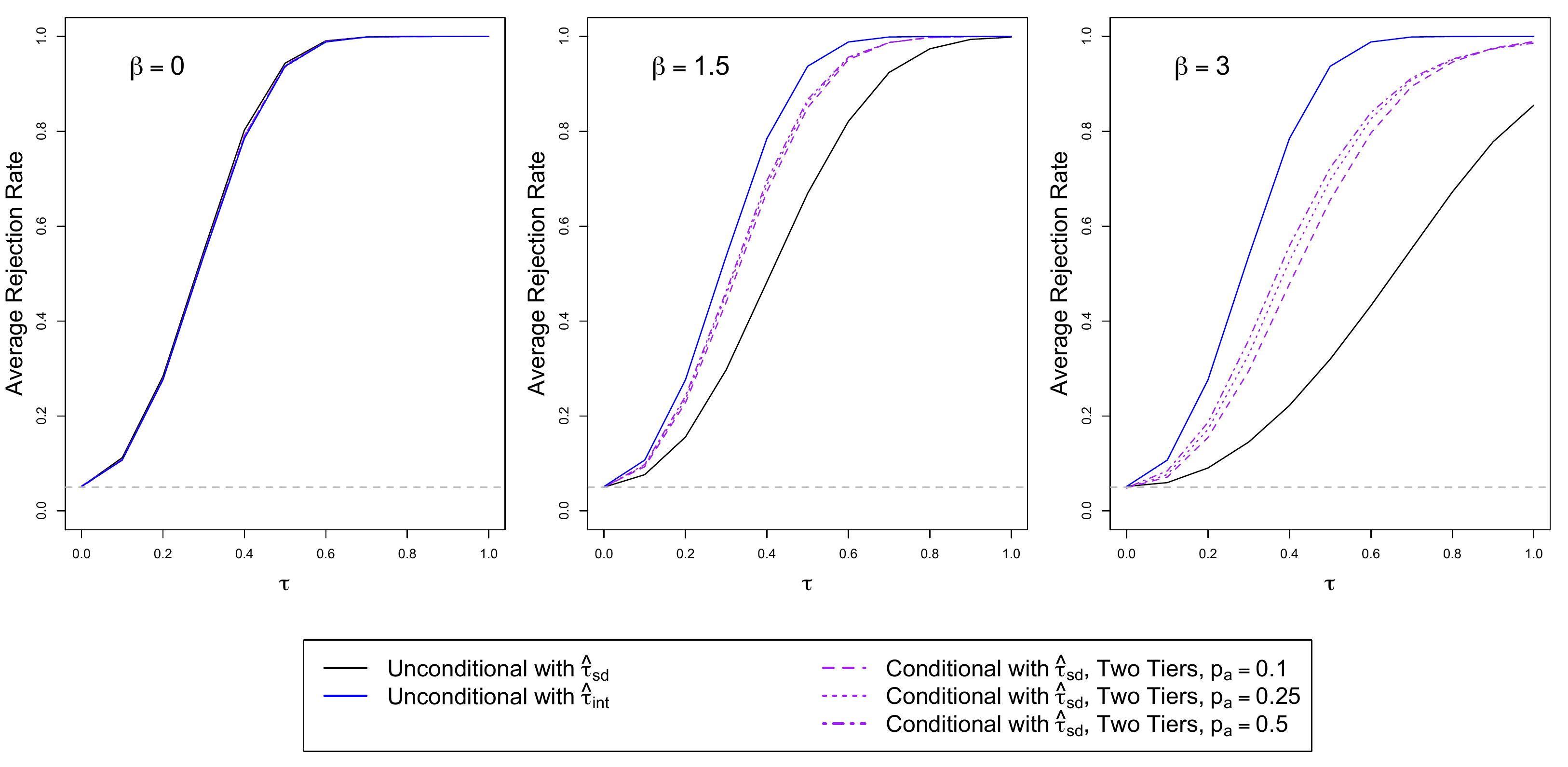}
    \caption{Power results of our conditional randomization test using two tiers.}
    \label{fig:powerPlotDifferentPaTwoTiers}
 \end{subfigure} 
 \caption{The rejection rate of the same tests discussed in Figure \ref{fig:powerPlotDifferentPa}, but for one or two tiers instead of four.}
 \label{fig:powerPlotDifferentPaOneTwoTiers}
  \end{figure}

Consider the same simulation setup as Section \ref{s:simulations}, where the potential outcomes for $N = 100$ units are generated using the model (\ref{eqn:potentialOutcomesModel}). In Section \ref{ss:unconditionalProperties}, we examined the power of our conditional randomization test for various acceptance probabilities for a fixed number of four tiers. Figure \ref{fig:powerPlotDifferentPaOneTwoTiers} shows the same results for one and two tiers, respectively. In other words, Figure \ref{fig:powerPlotDifferentPaOneTwoTiers} is analogous to Figure \ref{fig:powerPlotDifferentPa}, but for one or two tiers instead of four. The results are quite similar to those presented in Figure \ref{fig:powerPlotDifferentPa}: the power of our conditional randomization test increases as the acceptance probability decreases. Furthermore, by comparing Figures \ref{fig:powerPlotDifferentPa} and \ref{fig:powerPlotDifferentPaOneTwoTiers}, one can see that the additional benefit of decreasing the acceptance probability increases with the number of tiers. This emphasizes the benefit of conditioning on multiple measures of balance, rather than just a single measure.

\begin{figure}[H]
    \centering
    \begin{subfigure}[t]{\textwidth}
        \centering
        \includegraphics[scale=0.42]{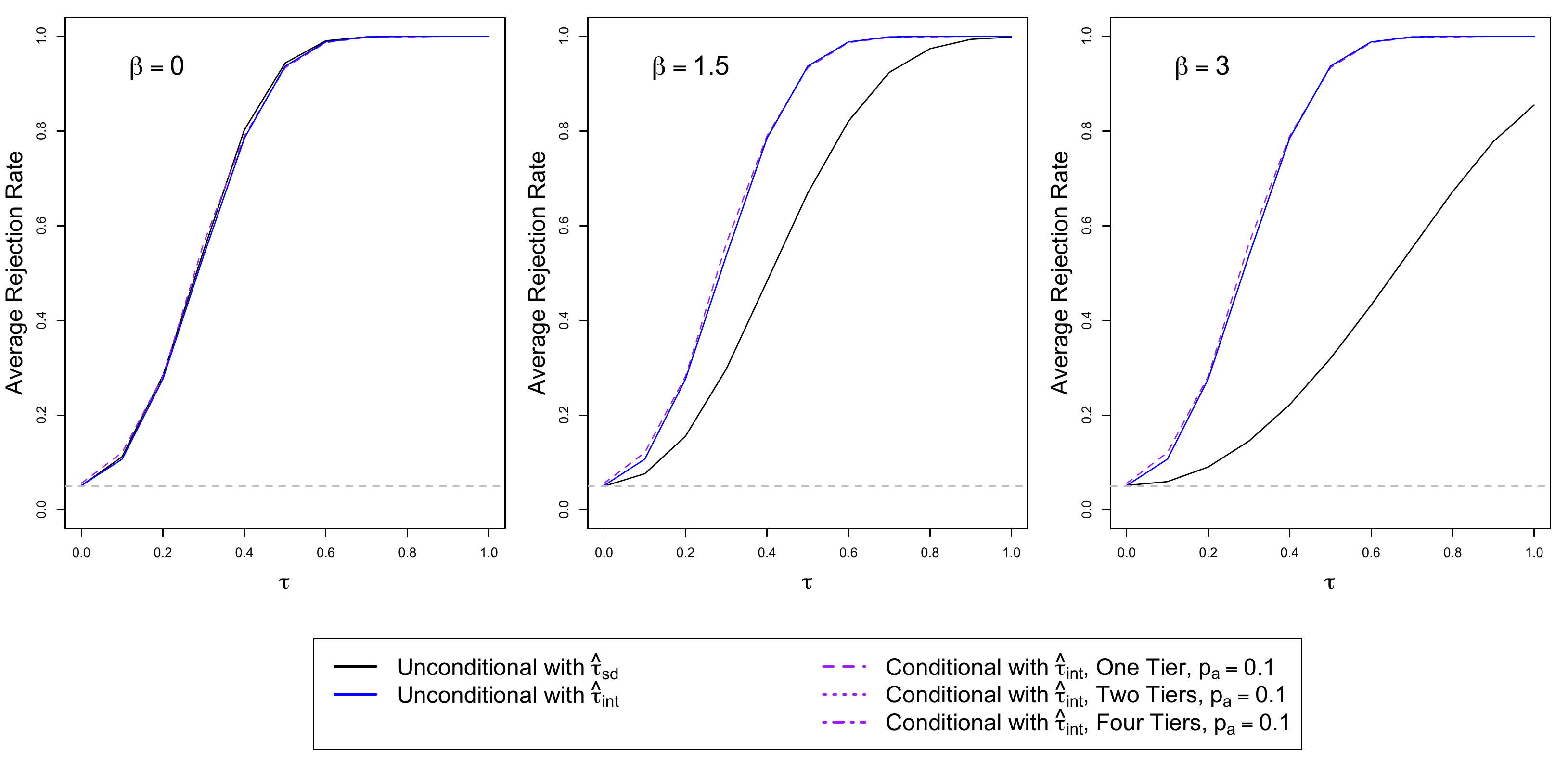}
        \caption{Conditional randomization tests using $\hat{\tau}_{int}$ for various tiers and a fixed acceptance probability.}
 	\label{fig:powerPlotRerandomizationWinston}
    \end{subfigure}%

    \begin{subfigure}[t]{\textwidth}
        \centering
        \includegraphics[scale=0.5]{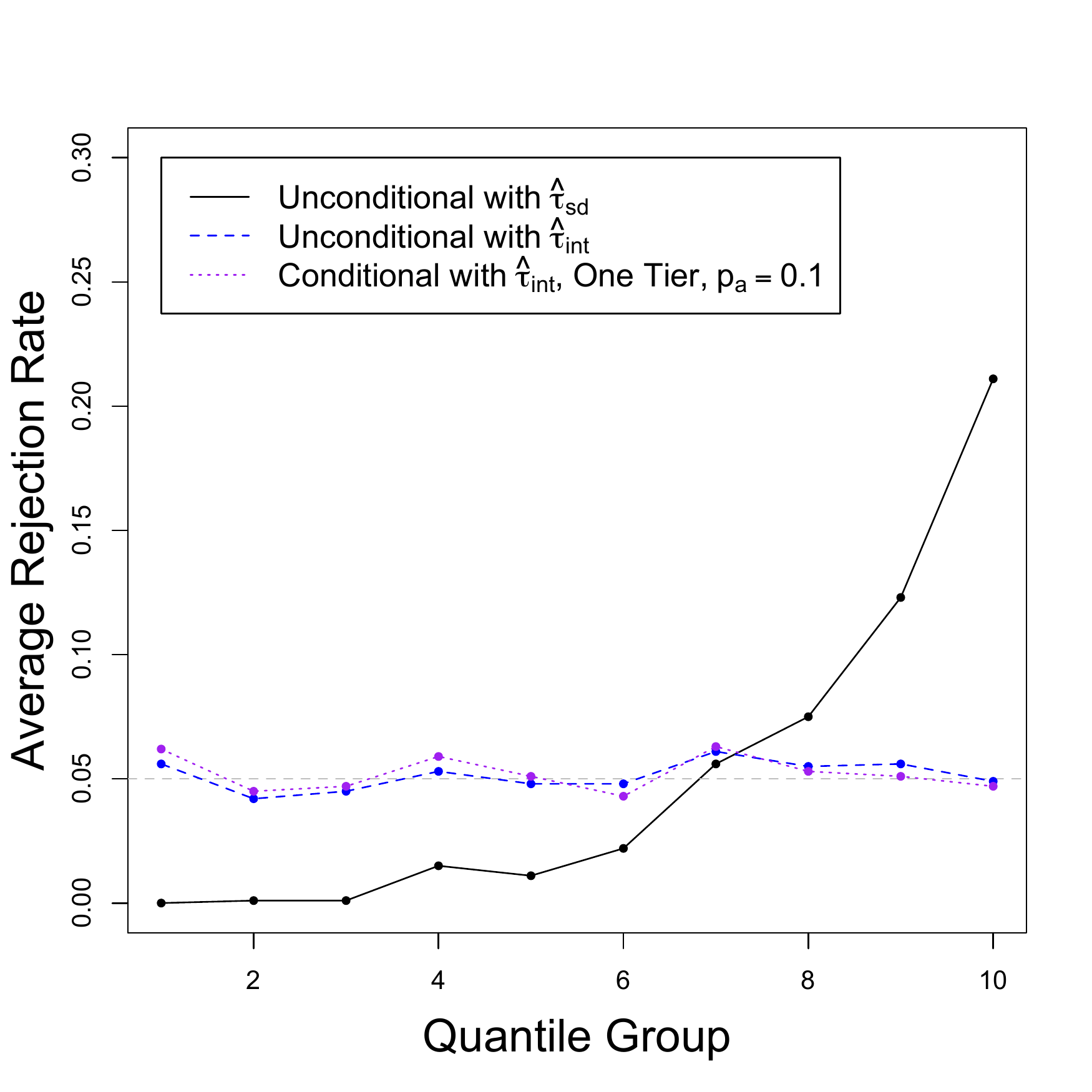}
	\caption{Average rejection rate of each randomization test when Fisher's Sharp Null Hypothesis is true. Rejection rates are shown within each quantile group of the Mahalanobis distance, such that each quantile group corresponds to 1,000 randomizations. Data were generated using (\ref{eqn:potentialOutcomesModel}) with $\tau = 0$ and $\beta = 3$, as in Section \ref{ss:conditionalProperties}.}
	\label{fig:conditionalPropertiesRerandomizationWinston}
    \end{subfigure}%
    \caption{The unconditional and conditional performance of our conditional randomization test using $\hat{\tau}_{int}$. Figure \ref{fig:powerPlotRerandomizationWinston} is analogous to Figure \ref{fig:powerPlotDifferentTiers}; Figure \ref{fig:conditionalPropertiesRerandomizationWinston} is analogous to Figure \ref{fig:conditionalProperties}.}
    \label{fig:rerandomizationWinstonUnconditionalConditionalProperties}
\end{figure}

Furthermore, in Section \ref{s:simulations} we focused on our conditional randomization test using the simple mean-difference test statistic $\hat{\tau}_{sd}$. Figure \ref{fig:rerandomizationWinstonUnconditionalConditionalProperties} presents the unconditional and conditional performance of our conditional randomization test using the regression-adjusted test statistic $\hat{\tau}_{int}$. In other words, Figures \ref{fig:powerPlotRerandomizationWinston} and \ref{fig:conditionalPropertiesRerandomizationWinston} are the same as Figures \ref{fig:powerPlot} and \ref{fig:conditionalProperties}, respectively, except we use $\hat{\tau}_{int}$ instead of $\hat{\tau}_{sd}$ for our conditional randomization test. We find that the power results for our conditional randomization test using $\hat{\tau}_{int}$ are essentially the same as those using $\hat{\tau}_{sd}$, and thus there does not appear to be an additional benefit of using a conditional randomization distribution for the randomization test if a model-adjusted test statistic is used (or vice versa).

\subsection{Simulation Results for Unbalanced Designs} \label{ss:unbalancedDesigns}

Consider the same simulation setup as Section \ref{s:simulations}, where the potential outcomes for $N = 100$ units are generated using the model (\ref{eqn:potentialOutcomesModel}). In Section \ref{s:simulations}, we considered balanced designs, where an equal number of units are assigned to treatment and control (i.e., $N_T = N_C = 50$). Here we consider an unbalanced design, where $N_T = 25$ and $N_C = 75$. Otherwise, the simulation setup discussed here is identical to the one discussed in Section \ref{s:simulations}. The results for this unbalanced design scenario are essentially identical to the results discussed in Section \ref{s:simulations}.

\begin{figure}[H]
	\centering
	\begin{subfigure}{\textwidth}
	\centering
 	\includegraphics[scale=0.5]{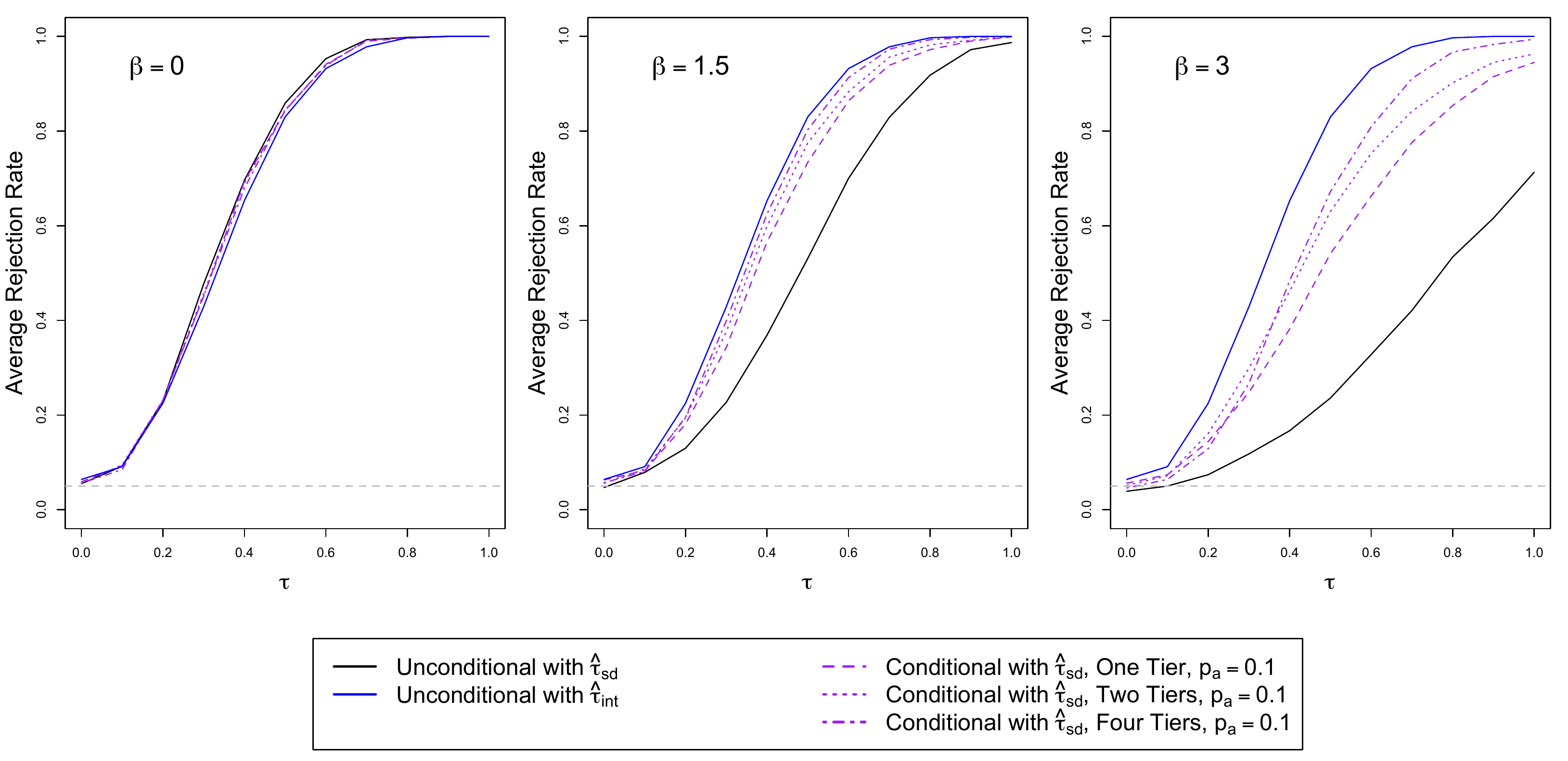}
 	\caption{For the conditional randomization test, different tiers and a fixed $p_a = 0.1$ acceptance probability.}
 	\label{fig:powerPlotDifferentTiersUnbalanced}
 	\end{subfigure}
 	\begin{subfigure}{\textwidth}
    \centering
    \includegraphics[scale=0.5]{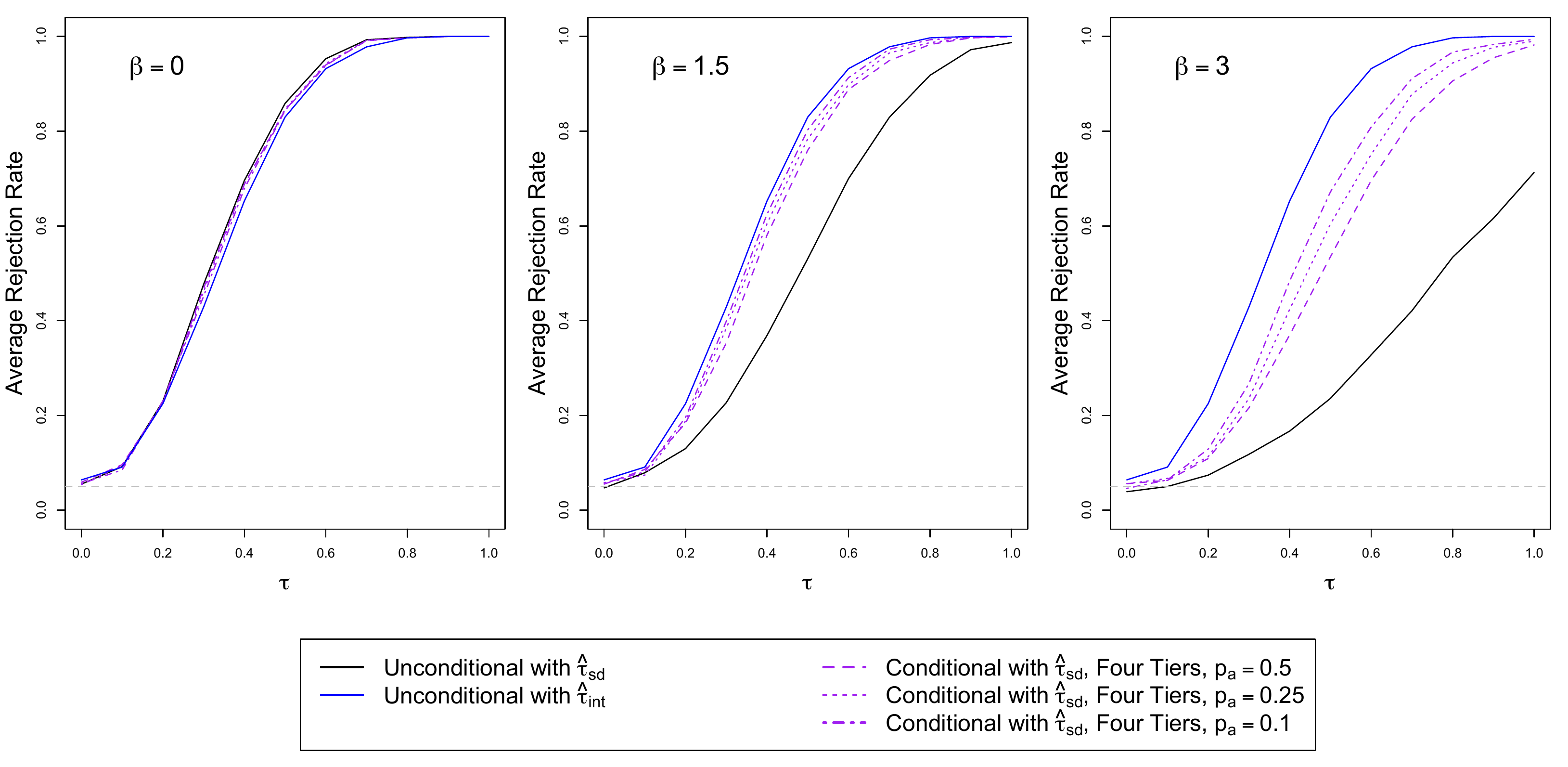}
    \caption{For the conditional randomization test, different acceptance probabilites and a fixed $T = 4$ tiers.}
    \label{fig:powerPlotDifferentPaUnbalanced}
 \end{subfigure} 
 \caption{The rejection rate of the same tests discussed in Figure \ref{fig:powerPlot}, but for an unbalanced design where $N_T = 25$ and $N_C = 75$ instead of a balanced design where $N_T = N_C = 50$.}
 \label{fig:powerPlotUnbalanced}
 \end{figure}

Figure \ref{fig:powerPlotUnbalanced} shows the power results of (1) the unconditional randomization test using $\hat{\tau}_{sd}$; (2) the unconditional randomization test using $\hat{\tau}_{int}$; and (3) our conditional randomization test using $\hat{\tau}_{sd}$. In other words, Figure \ref{fig:powerPlotUnbalanced} is analogous to Figure \ref{fig:powerPlot}, except the results are for an unbalanced design where $N_T = 25$ and $N_C = 75$ instead of a balanced design where $N_T = N_C = 50$. The power of all three tests are slightly lower for this case as compared to their power for the balanced design, but otherwise the results from Figure \ref{fig:powerPlotUnbalanced} are identical to those from Figure \ref{fig:powerPlot}: Our conditional randomization test is more powerful than the unconditional randomization test using $\hat{\tau}_{sd}$, and the results of our conditional randomization test approach those of the unconditional randomization test using $\hat{\tau}_{int}$ when the number of tiers increases or the acceptance probability $p_a$ decreases.

\begin{figure}[H]
    \centering
    \begin{subfigure}{\textwidth}
    \centering
    \includegraphics[scale=0.375]{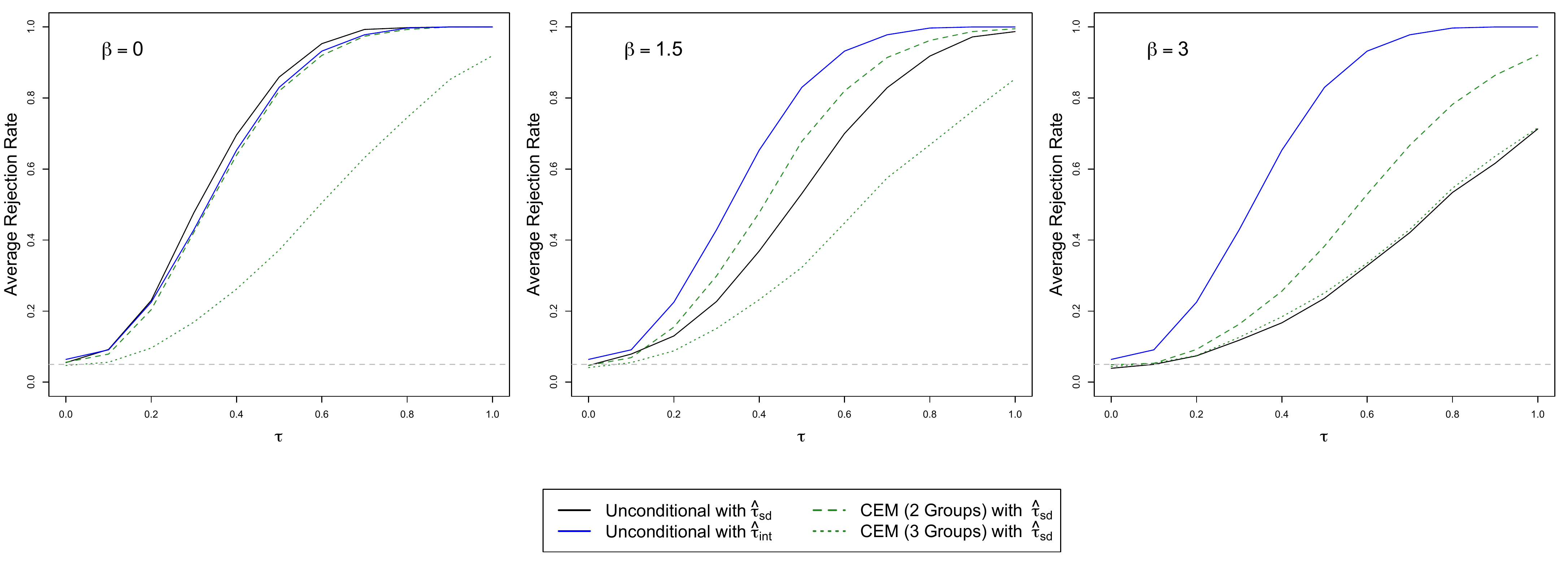}
    \caption{CEM (Prespecified Groups).}
    \label{fig:cemPowerPlot}
 \end{subfigure} 
  \begin{subfigure}{\textwidth}
    \centering
    \includegraphics[scale=0.375]{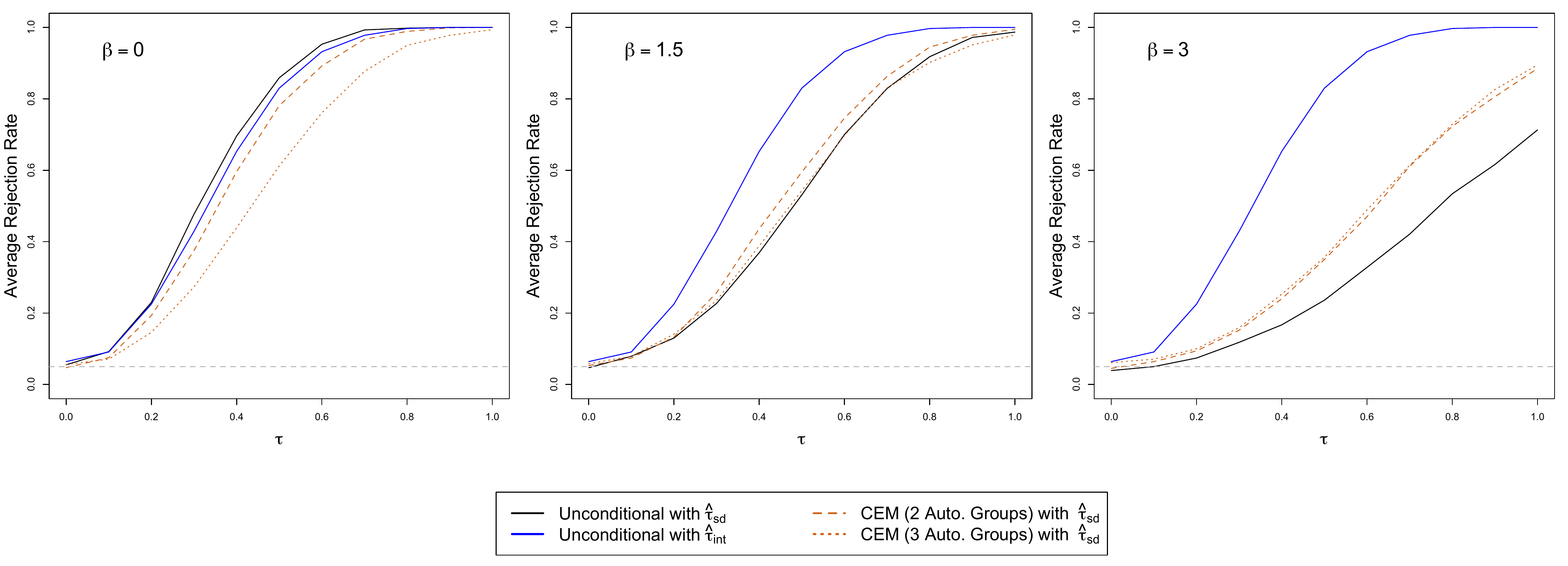}
    \caption{CEM (Automatic Groups).}
    \label{fig:cemAutoPowerPlot}
    \end{subfigure}
    \caption{Power results for the CEM (Prespecified Groups) and CEM (Automatic Groups) procedures for various number of groups for each covariate under the unbalanced design scenario where $N_T = 25$ and $N_C = 75$. Also shown are results for the unconditional tests using $\hat{\tau}_{sd}$ and $\hat{\tau}_{int}$, which are the same results from Figure \ref{fig:powerPlotUnbalanced}.}
    \label{fig:cemUnbalancedPowerPlots}
 \end{figure}

Meanwhile, Figure \ref{fig:cemUnbalancedPowerPlots} shows the power results of the CEM-based randomization tests discussed in Section \ref{s:simulations}. In other words, Figure \ref{fig:cemUnbalancedPowerPlots} is analogous to Figure \ref{fig:cemPowerPlots}, except the results are for the unbalanced design instead of the balanced design. For this unbalanced design scenario, we were only able to obtain results for $G = 2$ and $G = 3$ groups for the CEM-based tests, the reason being that there were less treated units in this unbalanced design, and thus less opportunities for CEM to find matches across many strata. This problem is the same as the issue that the CEM-based tests discard more and more units as the number of groups (or coarsened strata) $G$ increases, as discussed in Section \ref{s:simulations}. This again emphasizes the benefit of conditioning on forms of covariate balance that account for continuous covariates, instead of conditioning on a coarsened version of the covariate space. Otherwise, the results from Figure \ref{fig:cemUnbalancedPowerPlots} are identical to those from Figure \ref{fig:cemPowerPlots}: These CEM-based tests tend to be more powerful than the unconditional randomization test that uses $\hat{\tau}_{sd}$ but not as powerful as our conditional randomization test, and their power tends to decrease as $G$ increases.

\begin{figure}[H]
		\centering
	\includegraphics[scale=0.65]{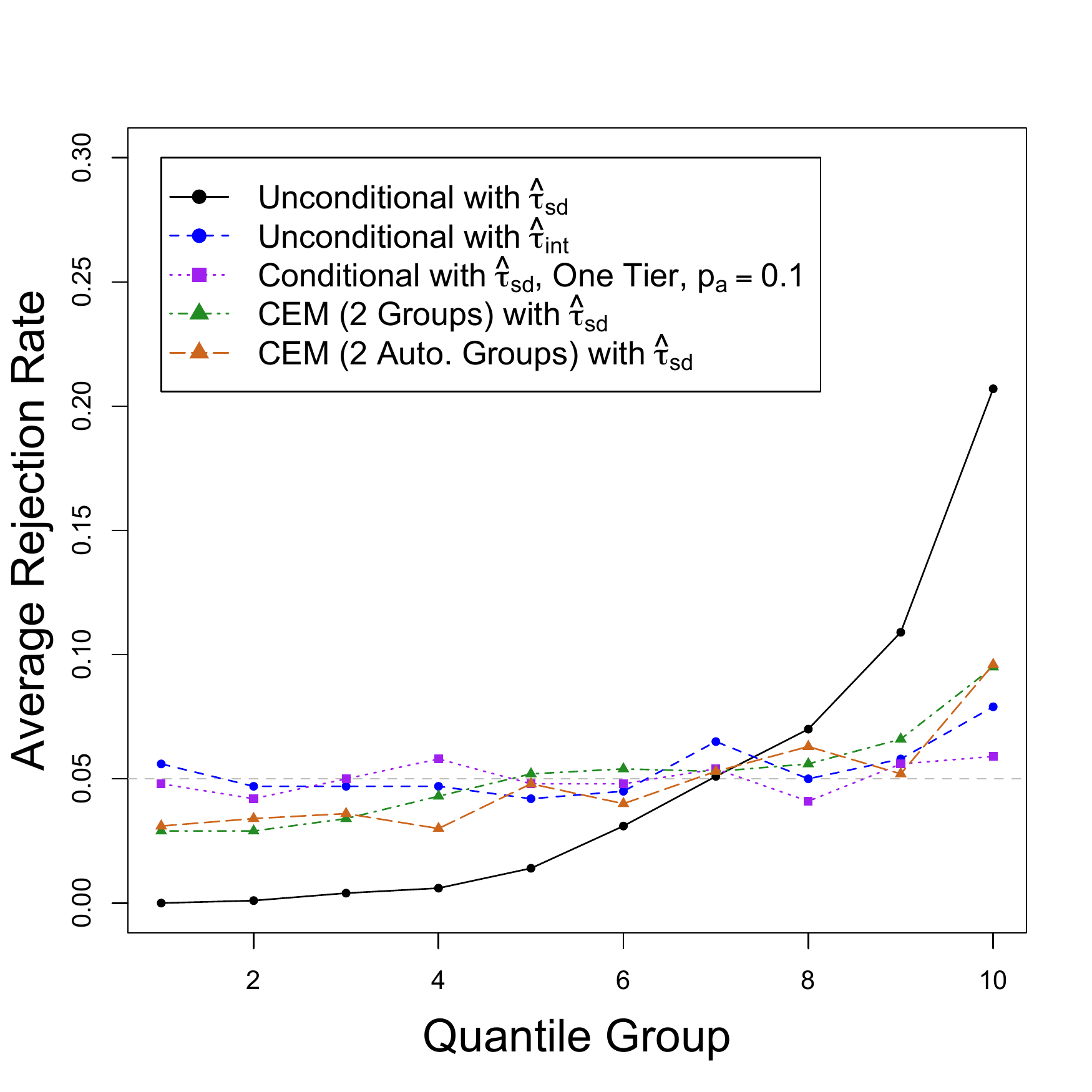}
	\caption{The rejection rate of the five randomization tests when Fisher's Sharp Null Hypothesis is true (i.e., $\tau = 0$) and $\beta = 3$ for the unbalanced design scenario. Rejection rates are shown within each quantile group of the Mahalanobis distance, such that each quantile group corresponds to 1,000 randomizations.}
	\label{fig:conditionalPropertiesUnbalanced}
\end{figure}

Similar to Section \ref{ss:conditionalProperties}, we also examined the conditional performance of these randomization tests for this unbalanced design scenario. After the potential outcomes were generated from (\ref{eqn:potentialOutcomesModel}) for $\tau = 0$ and $\beta = 3$, we simulated 10,000 randomizations (where $N_T = 25$ and $N_C = 75$) and computed the Mahalanobis distance for each randomization. Then, we divided these randomizations into 10 groups according to the 10 quantiles of the 10,000 Mahalanobis distances. Figure \ref{fig:conditionalPropertiesUnbalanced} shows the rejection rate of each of the five randomization tests for each quantile group of the Mahalanobis distance. In other words, Figure \ref{fig:conditionalPropertiesUnbalanced} is analogous to Figure \ref{fig:conditionalProperties}, except for the unbalanced design instead of the balanced design. The results are again largely the same as those presented in Section \ref{ss:conditionalProperties}: The unconditional randomization test using $\hat{\tau}_{int}$ and the conditional randomization test using $\hat{\tau}_{sd}$ are conditionally valid across quantile groups, while the unconditional randomization test using $\hat{\tau}_{sd}$ is not conditionally valid and its rejection rate is monotonically increasing in covariate imbalance. Meanwhile, similar to Section \ref{ss:conditionalProperties}, the false rejection rate for the CEM-based tests also appears to be monotonically increasing in covariate imbalance according to the Mahalanobis distance, but to a much less severe degree, suggesting that these tests are approximately conditionally valid.

\subsection{Simulation Results for Alternative Data-Generating Linear Models} \label{ss:alternativeLinearModels}

In Section \ref{s:simulations}, the potential outcomes were generated using the linear model (\ref{eqn:potentialOutcomesModel}) where all the covariates had positive effects on the outcomes, were unrelated to the treatment effect, and were normally distributed. Here we consider alternative linear models for the potential outcomes and compare power results for the unconditional randomization tests using $\hat{\tau}_{sd}$ and $\hat{\tau}_{int}$ as well as our conditional randomization test using $\hat{\tau}_{sd}$ for these alternative models. We examine the performance of the randomization tests under each of the following models:
\begin{itemize}
\item \textbf{Positive/Negative Covariate Effects}
\begin{equation}
	\begin{aligned}
		&Y_i(0) | X_i = \beta(-0.1X_{i1} + 0.2X_{i2} + 0.3X_{i3} - 0.4X_{i4}) + \epsilon_i, \hspace{0.1 in} i=1,\dots,100 \\
		&Y_i(1) = Y_i(0) + \tau \label{eqn:potentialOutcomesModelPositiveNegative}
	\end{aligned}
	\end{equation}
	where $(X_{i1}, X_{i2},X_{i3}, X_{i4},\epsilon_i) \stackrel{iid}{\sim} \mathcal{N}_5(0,I_5)$.
\item \textbf{Heterogeneous Treatment Effects}
\begin{equation}
	\begin{aligned}
		&Y_i(0) | X_i = \beta(0.1X_{i1} + 0.2X_{i2} + 0.3X_{i3} + 0.4X_{i4}) + \epsilon_i, \hspace{0.1 in} i=1,\dots,100 \\
		&Y_i(1) = Y_i(0) + \tau + \sigma_{\tau}Y_i(0) \label{eqn:potentialOutcomesModelHeterogeneous}
	\end{aligned}
	\end{equation}
	where $(X_{i1}, X_{i2},X_{i3}, X_{i4},\epsilon_i) \stackrel{iid}{\sim} \mathcal{N}_5(0,I_5)$. Following \cite{ding2016randomization}, we set $\sigma_{\tau} = 0.5$ to induce strong treatment effect heterogeneity.
\item \textbf{Different Covariate Distributions}
\begin{equation}
	\begin{aligned}
		&Y_i(0) | X_i = \beta(0.1X_{i1} + 0.2X_{i2} + 0.3X_{i3} + 0.4X_{i4}) + \epsilon_i, \hspace{0.1 in} i=1,\dots,100 \\
		&Y_i(1) = Y_i(0) + \tau \label{eqn:potentialOutcomesModelDifferentDistributions}
	\end{aligned}
	\end{equation}
    where $X_{i1} \sim N(0, 1), X_{i2} \sim N(X_{i1}, 1), X_{i3} \sim \text{Pois}(5), X_{i4} \sim \text{Bern}(0.2)$, and $\epsilon_i \sim N(0,1)$.
\end{itemize}
Similar to Section \ref{s:simulations}, the parameters $\beta$ and $\tau$ take on values $\beta \in \{0, 1.5, 3\}$ and $\tau \in \{0, 0.1, \dots 1\}$ across simulations for the above models.

\begin{figure}[H]
    \centering
    \begin{subfigure}[t]{\textwidth}
        \centering
        \includegraphics[scale=0.42]{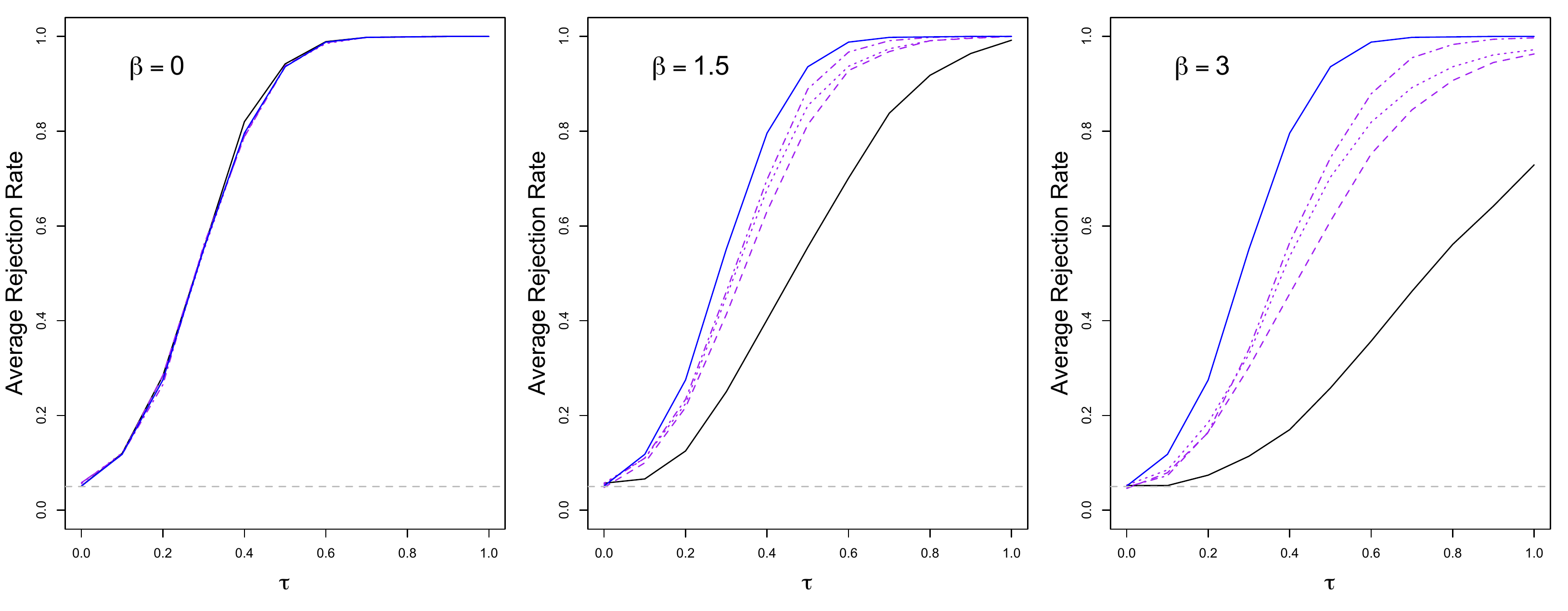}
        \caption{Potential outcomes generated from the Positive/Negative Covariate Effects model (\ref{eqn:potentialOutcomesModelPositiveNegative}).}
    \end{subfigure}%
    
    \begin{subfigure}[t]{\textwidth}
        \centering
        \includegraphics[scale=0.42]{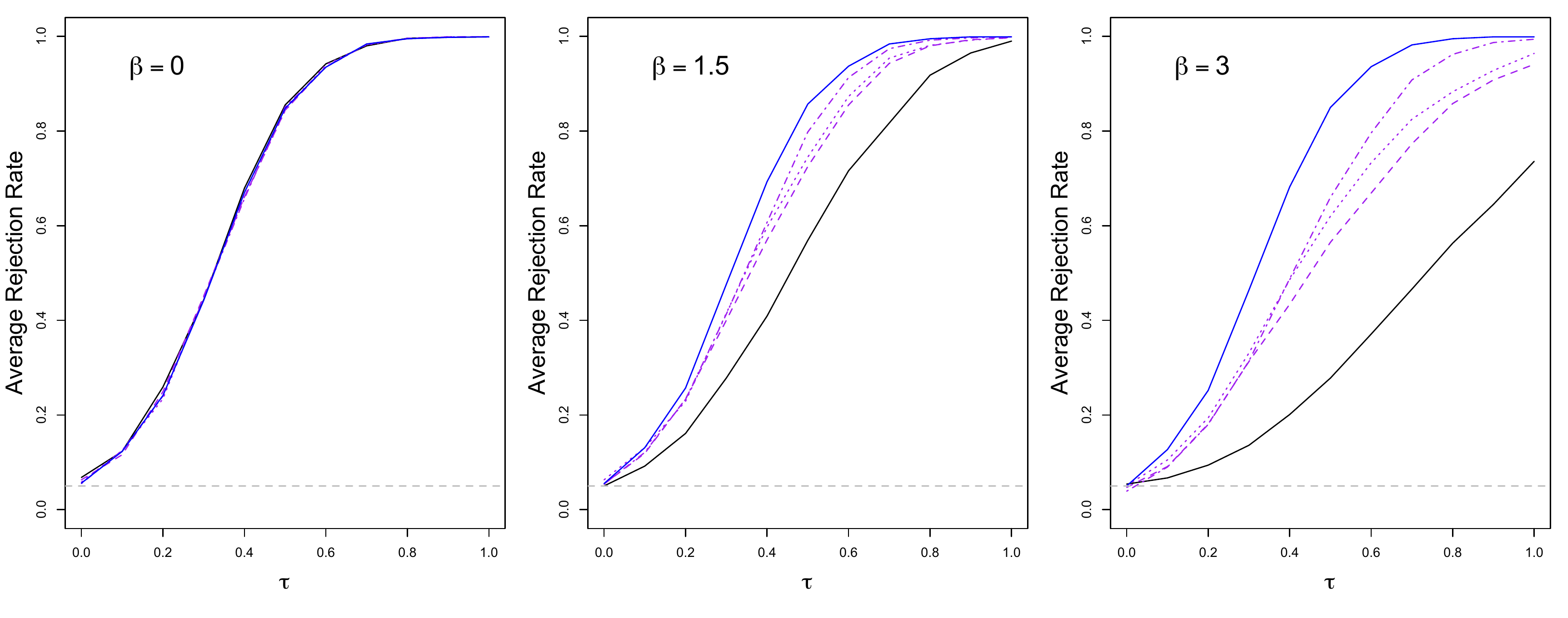}
        \caption{Potential outcomes generated from the Heterogeneous Treatment Effects model (\ref{eqn:potentialOutcomesModelHeterogeneous}).}
    \end{subfigure}

    \begin{subfigure}[t]{\textwidth}
        \centering
        \includegraphics[scale=0.42]{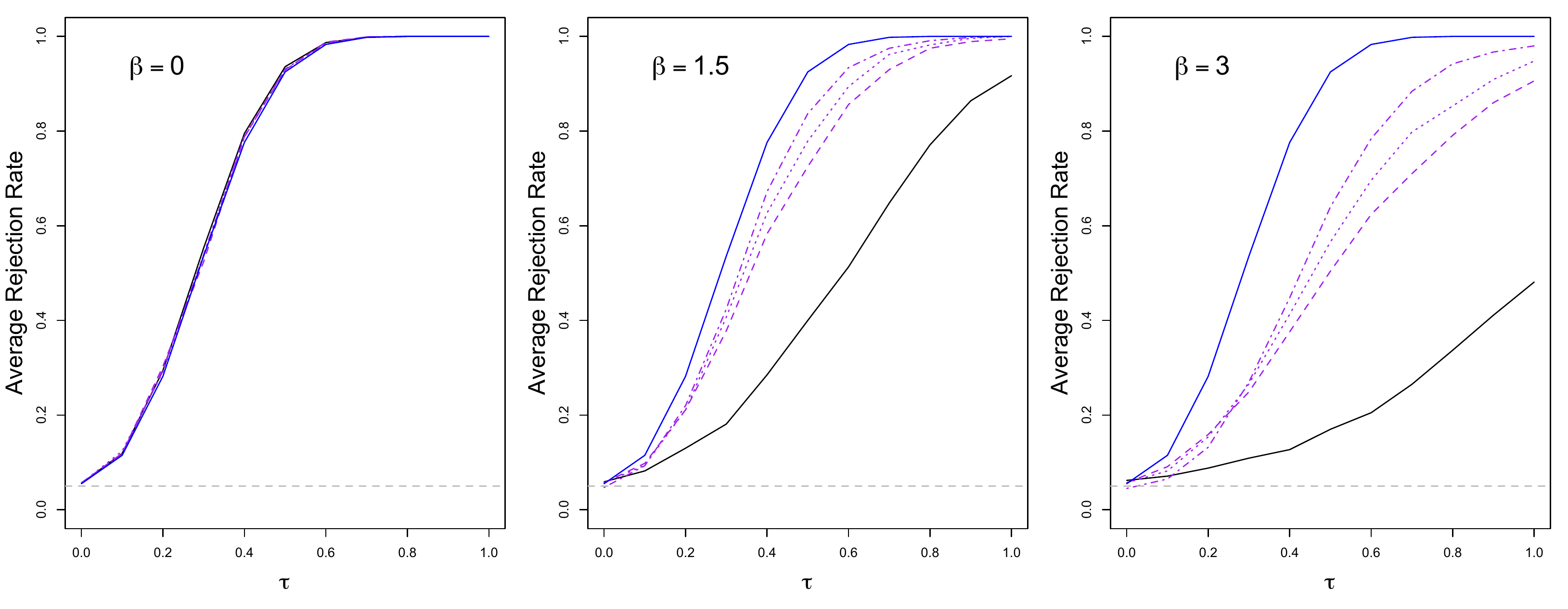}
        \caption{Potential outcomes generated from the Different Covariate Distributions model (\ref{eqn:potentialOutcomesModelDifferentDistributions}).}
    \end{subfigure}

        \includegraphics[scale=0.5]{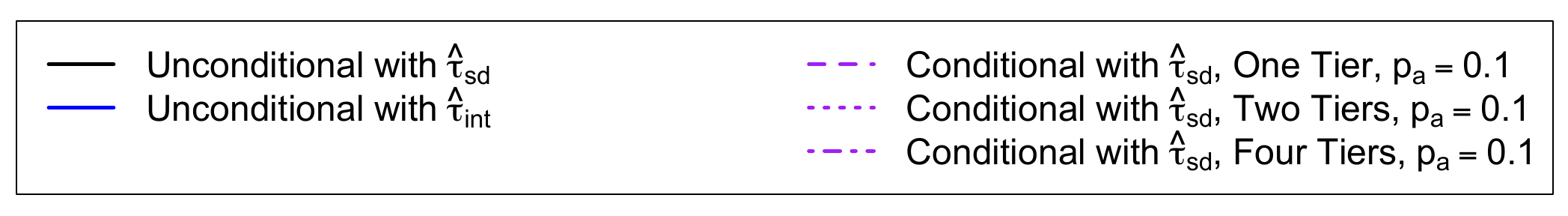}

    \caption{Average rejection rate for the unconditional randomization tests using $\hat{\tau}_{sd}$ and $\hat{\tau}_{int}$ as well as our conditional randomization test using $\hat{\tau}_{sd}$ for various tiers and a fixed acceptance probability, where the potential outcomes were generated from the Positive/Negative Covariate Effects model (\ref{eqn:potentialOutcomesModelPositiveNegative}), Heterogeneous Treatment Effects model (\ref{eqn:potentialOutcomesModelHeterogeneous}), or Different Covariate Distributions model (\ref{eqn:potentialOutcomesModelDifferentDistributions}).}
    \label{fig:powerPlotPositiveNegativeHeterogeneousDifferentDistributions}
\end{figure}

Figure \ref{fig:powerPlotPositiveNegativeHeterogeneousDifferentDistributions} shows the power results of the randomization tests when the potential outcomes were generated from the above models. Figure \ref{fig:powerPlotPositiveNegativeHeterogeneousDifferentDistributions} is analogous to Figure \ref{fig:powerPlot}, except the potential outcomes were generated from models (\ref{eqn:potentialOutcomesModelPositiveNegative}), (\ref{eqn:potentialOutcomesModelHeterogeneous}), or (\ref{eqn:potentialOutcomesModelDifferentDistributions}) instead of model (\ref{eqn:potentialOutcomesModel}) used in Section \ref{s:simulations}. The results are largely the same: The conditional randomization test is more powerful than the unconditional randomization test that uses the unadjusted test statistic $\hat{\tau}_{sd}$; furthermore, as the number of tiers increases, the conditional randomization test approaches the unconditional randomization test that uses the regression-adjusted test statistic.

\begin{figure}[H]
    \centering
    \begin{subfigure}[t]{0.32\textwidth}
        \centering
        \includegraphics[scale=0.32]{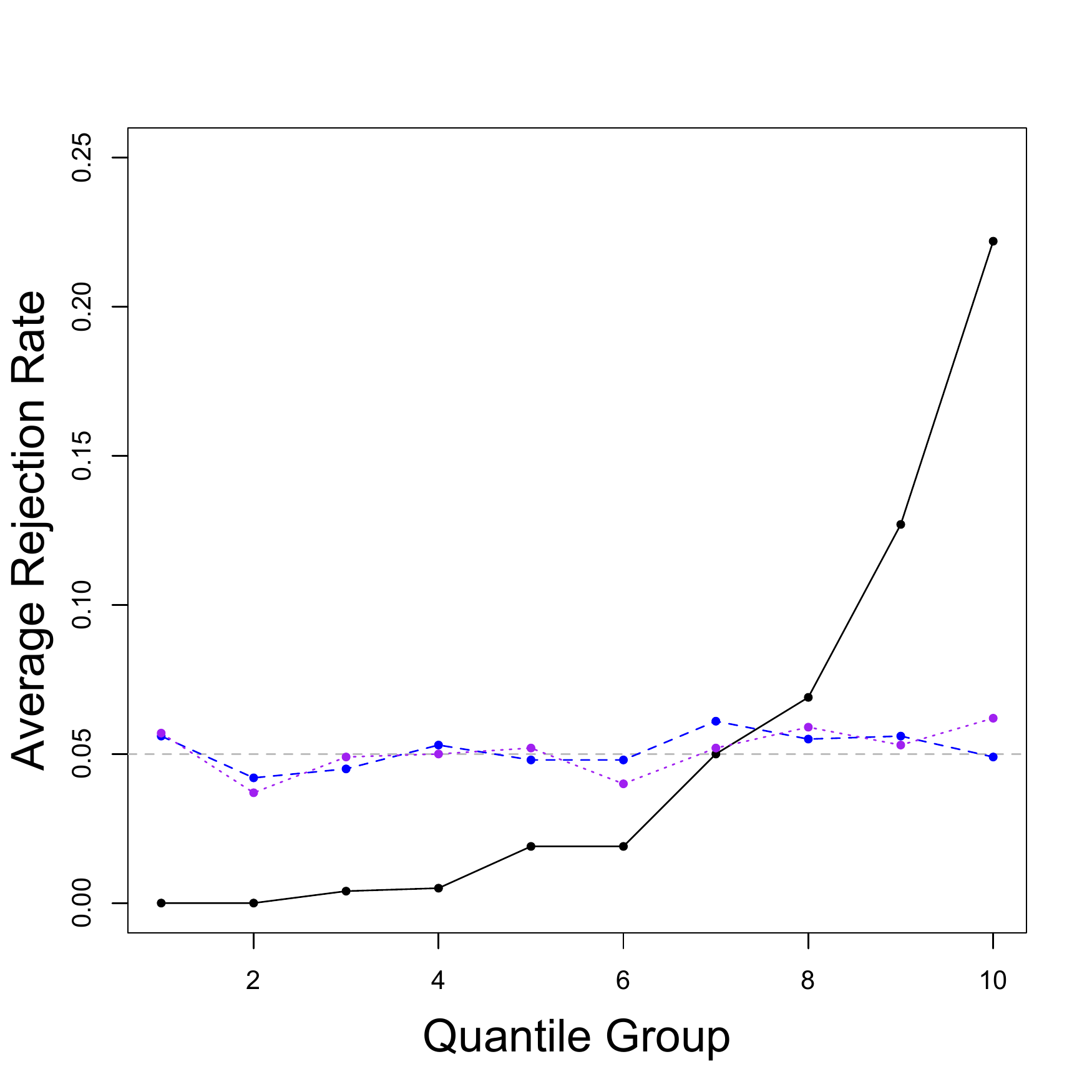}
        \caption{Positive/Negative Covariate Effects model (\ref{eqn:potentialOutcomesModelPositiveNegative}).}
    \end{subfigure}%
    ~
    \begin{subfigure}[t]{0.32\textwidth}
        \centering
        \includegraphics[scale=0.32]{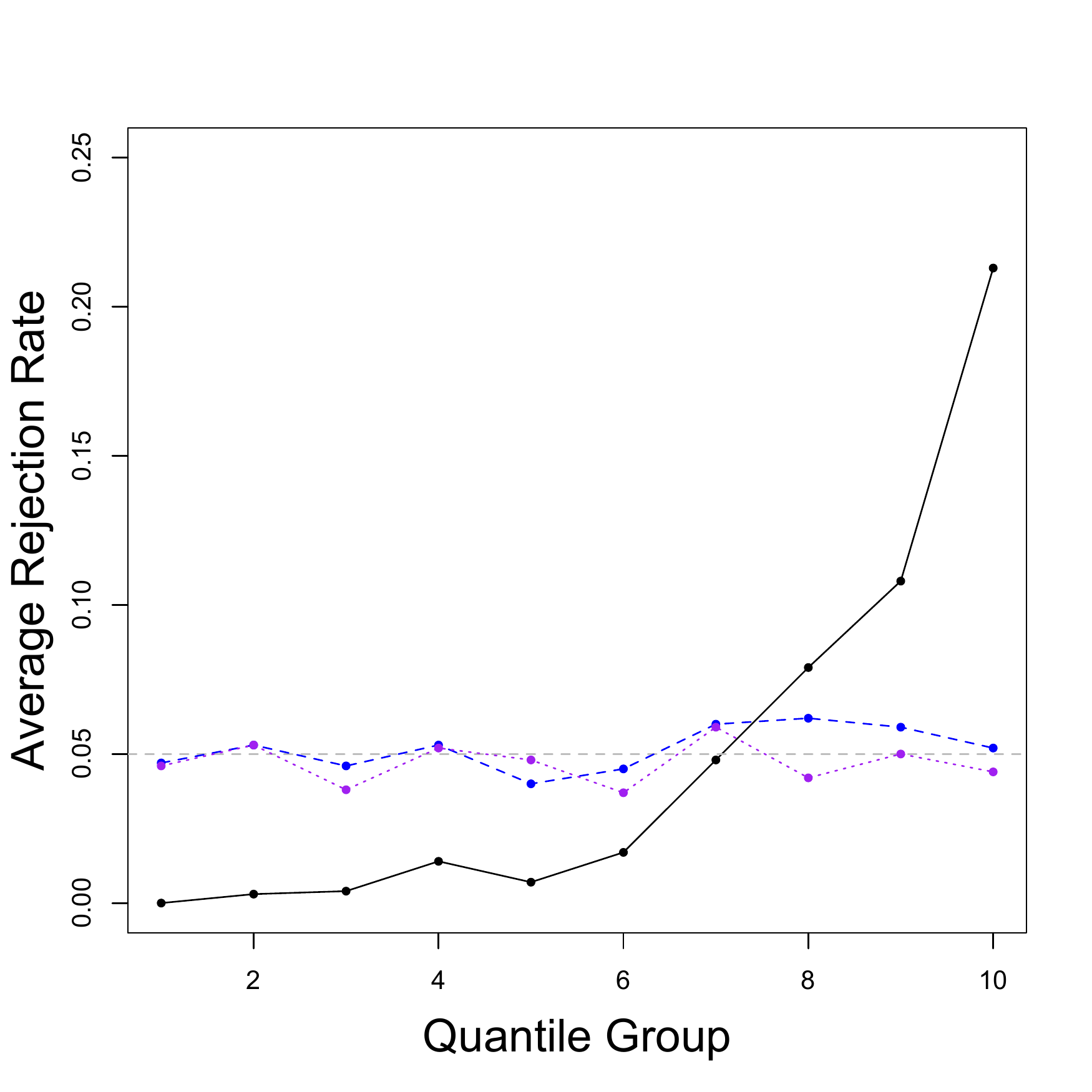}
        \caption{Heterogeneous Treatment Effects model (\ref{eqn:potentialOutcomesModelHeterogeneous}).}
    \end{subfigure}
    ~
    \begin{subfigure}[t]{0.32\textwidth}
        \centering
        \includegraphics[scale=0.32]{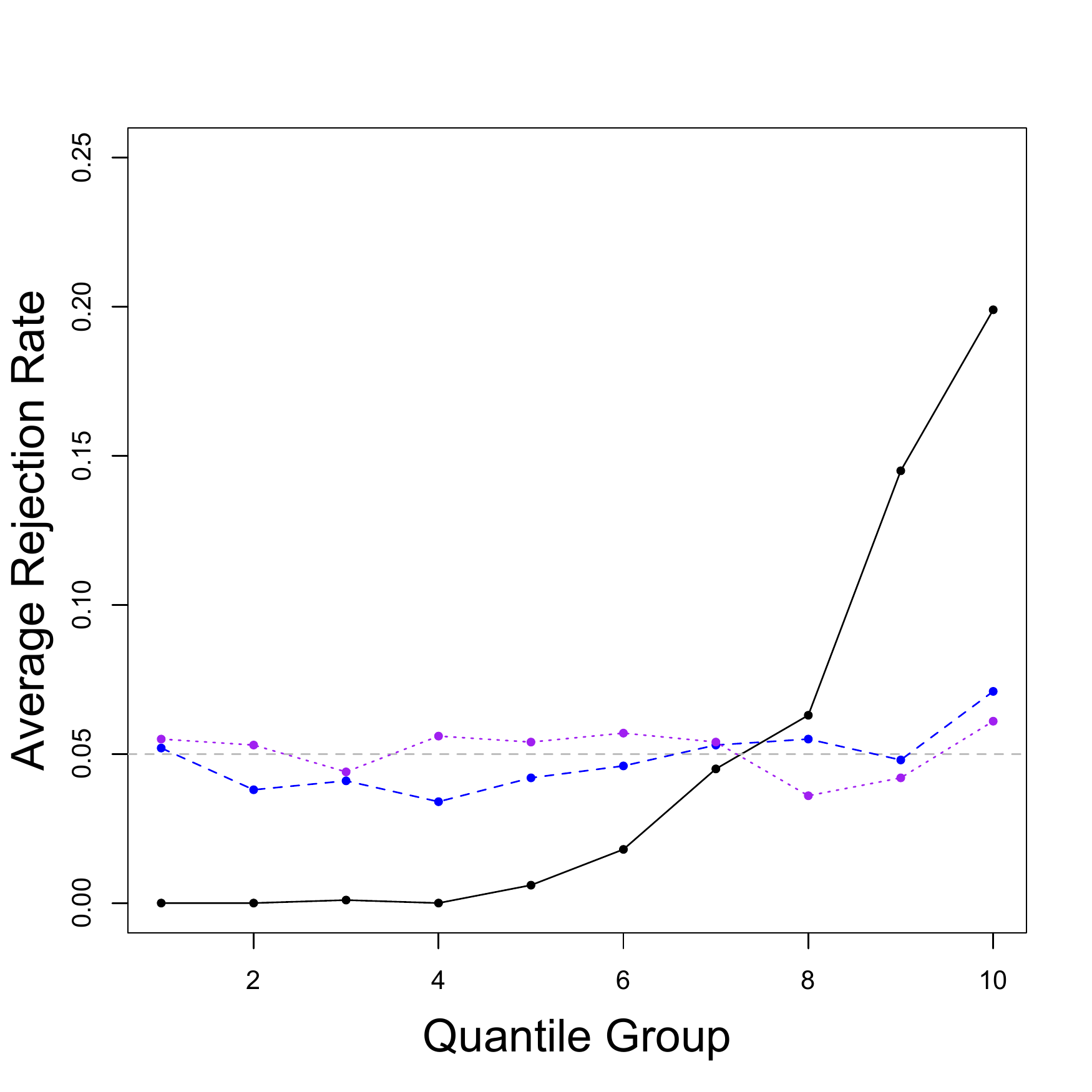}
        \caption{Different Covariate Distributions model (\ref{eqn:potentialOutcomesModelDifferentDistributions}).}
    \end{subfigure}

    \includegraphics[scale=0.5]{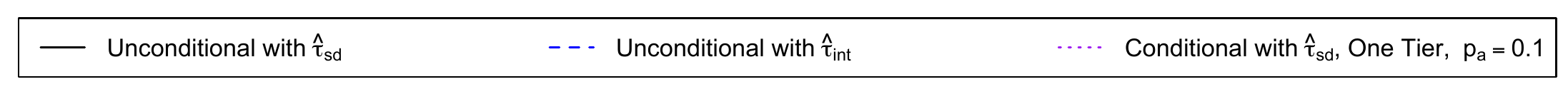}

    \caption{The rejection rate of the randomization tests within each quantile group of the Mahalanobis distance when the potential outcomes were generated from the Positive/Negative Covariate Effects model (\ref{eqn:potentialOutcomesModelPositiveNegative}), Heterogeneous Treatment Effects model (\ref{eqn:potentialOutcomesModelHeterogeneous}), or Different Covariate Distributions model (\ref{eqn:potentialOutcomesModelDifferentDistributions}).}
    \label{fig:conditionalPropertiesPositiveNegativeHeterogeneousDifferentDistributions}
\end{figure}

Similar to Section \ref{ss:conditionalProperties}, we also examined the conditional performance of the randomization tests when the potential outcomes were generated from the above models. After the potential outcomes were generated for $\tau = 0$ and $\beta = 3$ for each of the three models, we simulated 10,000 randomizations and computed the Mahalanobis distance for each randomization. Then, we divided these randomizations into 10 groups according to the 10 quantiles of the 10,000 Mahalanobis distances. Figure \ref{fig:conditionalPropertiesPositiveNegativeHeterogeneousDifferentDistributions} shows the rejection rate of each randomization test for each quantile group for each of the three potential outcome models. Figure \ref{fig:conditionalPropertiesPositiveNegativeHeterogeneousDifferentDistributions} is analogous to Figure \ref{fig:conditionalProperties}, except the potential outcomes were generated from models (\ref{eqn:potentialOutcomesModelPositiveNegative}), (\ref{eqn:potentialOutcomesModelHeterogeneous}), or (\ref{eqn:potentialOutcomesModelDifferentDistributions}) instead of model (\ref{eqn:potentialOutcomesModel}). The results are again largely the same as those presented in Section \ref{ss:conditionalProperties}: The unconditional randomization test using $\hat{\tau}_{int}$ and the conditional randomization test using $\hat{\tau}_{sd}$ are conditionally valid across quantile groups, while the unconditional randomization test using $\hat{\tau}_{sd}$ is not conditionally valid and its rejection rate is monotonically increasing in covariate imbalance. In short, Figures \ref{fig:powerPlotPositiveNegativeHeterogeneousDifferentDistributions} and \ref{fig:conditionalPropertiesPositiveNegativeHeterogeneousDifferentDistributions} suggest that the results found in Section \ref{s:simulations} hold across many data-generating processes.

\subsection{Simulation Results for Misspecified Linear Models} \label{ss:misspecifiedModels}

 In the simulation study discussed in Section \ref{s:simulations}, the potential outcomes were generated from the linear model (\ref{eqn:potentialOutcomesModel}). We considered using the test statistic $\hat{\tau}_{int}$, which is defined as the estimated coefficient for $W_i$ from the linear regression of $Y_i$ on $W_i$, $\mathbf{x}_i$, and $W_i(\mathbf{x}_i - \overline{\mathbf{X}})$. Thus, $\hat{\tau}_{int}$ is a correctly specified model in the simulation setup presented in Section \ref{s:simulations}. We now consider cases when $\hat{\tau}_{int}$ is still defined as in Section \ref{s:simulations} but the potential outcomes are generated from a nonlinear model, making the model $\hat{\tau}_{int}$ assumes misspecified. Consider $N = 100$ units whose potential outcomes are generated from one of the following models:
\begin{itemize}
\item \textbf{Model with Moderate Correlation}
\begin{equation}
	\begin{aligned}
		&Y_i(0) | X_i = \beta \left(0.1X_{i1}^2 + 0.2X_{i2} + 0.3X_{i3}^2 + 0.4X_{i4} \right) + \epsilon_i, \hspace{0.1 in} i=1,\dots,100 \\
		&Y_i(1) = Y_i(0) + \tau \label{eqn:potentialOutcomesModelMisspecifiedModerateLinear}
	\end{aligned}
	\end{equation}
	where $(X_{i1}, X_{i2},X_{i3}, X_{i4},\epsilon_i) \stackrel{iid}{\sim} \mathcal{N}_5(0,I_5)$.
\item \textbf{Model with No Correlation}
	\begin{equation}
	\begin{aligned}
		&Y_i(0) | X_i = \beta \left(0.1 \sqrt{|X_{i1}|} + 0.2X_{i2}^2 + 0.3 \sqrt{|X_{i3}|} + 0.4X_{i4}^2 \right) + \epsilon_i, \hspace{0.1 in} i=1,\dots,100 \\
		&Y_i(1) = Y_i(0) + \tau \label{eqn:potentialOutcomesModelMisspecifiedWeakLinear}
	\end{aligned}
	\end{equation}
	where $(X_{i1}, X_{i2},X_{i3}, X_{i4},\epsilon_i) \stackrel{iid}{\sim} \mathcal{N}_5(0,I_5)$.
\end{itemize}
Similar to Section \ref{s:simulations}, the parameters $\beta$ and $\tau$ take on values $\beta \in \{0, 1.5, 3\}$ and $\tau \in \{0, 0.1, \dots 1\}$ across simulations for the above models.

In the first model, there is a moderate correlation between the raw covariates and the potential outcomes: For the specific set of potential outcomes generated from (\ref{eqn:potentialOutcomesModelMisspecifiedModerateLinear}) with $\beta = 3$ for the simulation, the empirical $R^2$ between $\mathbf{Y}(0)$ and $(\mathbf{X}_1, \mathbf{X}_2, \mathbf{X}_3, \mathbf{X}_4)$ was 0.33. Meanwhile, in the second model, there is no correlation between the raw covariates and the potential outcomes: For the specific set of potential outcomes generated from (\ref{eqn:potentialOutcomesModelMisspecifiedWeakLinear}) with $\beta = 3$ for the simulation, the empirical $R^2$ was only 0.075.  These cases differ from the case discussed in Section \ref{s:simulations}, where the empricial $R^2$ was 0.82 and thus there was a strong correlation between the raw covariates and the potential outcomes.

\begin{figure}[H]
	\centering
	\begin{subfigure}[t]{\textwidth}
	\centering
 	\includegraphics[scale=0.5]{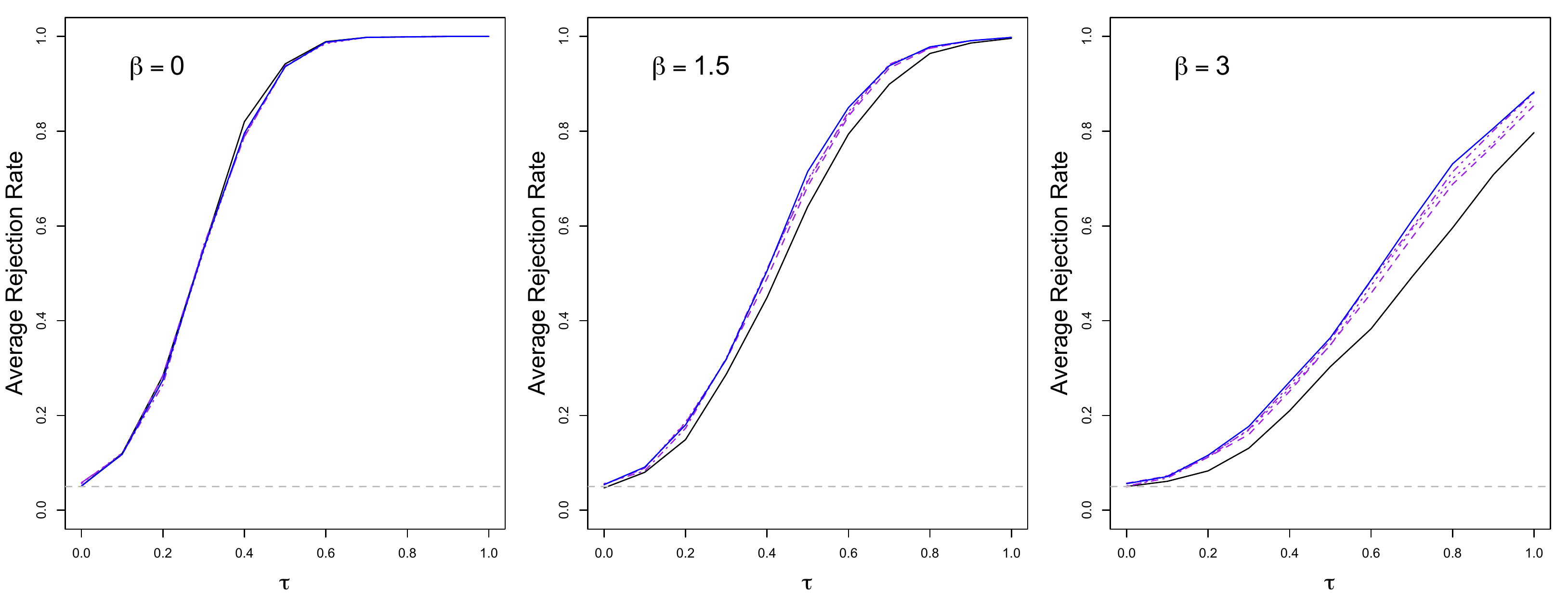}
 	\caption{Potential outcomes generated from the Moderate Correlation model (\ref{eqn:potentialOutcomesModelMisspecifiedModerateLinear}).}
 	\label{fig:powerPlotMisspecifiedModerateLinear}
 	\end{subfigure}

 	\begin{subfigure}[t]{\textwidth}
	\centering
 	\includegraphics[scale=0.5]{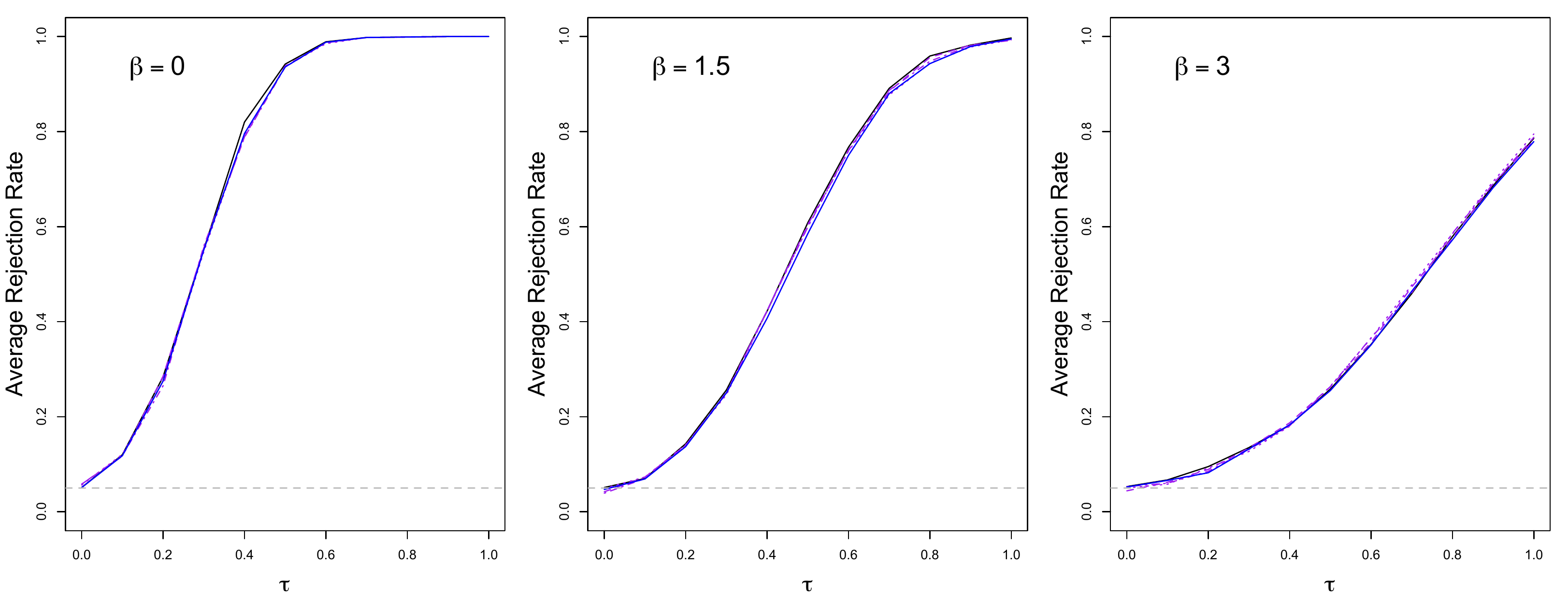}
 	\caption{Potential outcomes generated from the No Correlation model (\ref{eqn:potentialOutcomesModelMisspecifiedWeakLinear}).}
 	\label{fig:powerPlotMisspecifiedWeakLinear}
 	\end{subfigure}

 	\includegraphics[scale=0.5]{positiveNegativeHeterogeneousDifferentDistributionsLegend.pdf}
 	\caption{Average rejection rate of the unconditional randomization tests using $\hat{\tau}_{sd}$ and $\hat{\tau}_{int}$ as well as our conditional randomization test when the potential outcomes were generated from the Moderate Correlation model (\ref{eqn:potentialOutcomesModelMisspecifiedModerateLinear}) or the No Correlation model (\ref{eqn:potentialOutcomesModelMisspecifiedWeakLinear}).}
 	\label{fig:powerPlotMisspecifiedModels}
 \end{figure}

Figure \ref{fig:powerPlotMisspecifiedModels} shows the power results of the randomization tests when the potential outcomes were generated from the above models. The results for the Moderate Correlation case are similar to those presented in Section \ref{s:simulations}: The conditional randomization test is more powerful than the unconditional randomization test that uses $\hat{\tau}_{sd}$; furthermore, as the number of tiers increases, the conditional randomization test approaches the unconditional randomization test that uses $\hat{\tau}_{int}$. Meanwhile, for the No Correlation case, the power of all the tests appear to be essentially equivalent. These results suggest that there is a benefit of using our conditional randomization test or the unconditional randomization test with a regression-adjusted test statistic if there is a correlation between the covariates and the potential outcomes.

\begin{figure}[H]
	\centering
	\begin{subfigure}[t]{0.45\textwidth}
	\centering
	\includegraphics[scale=0.4]{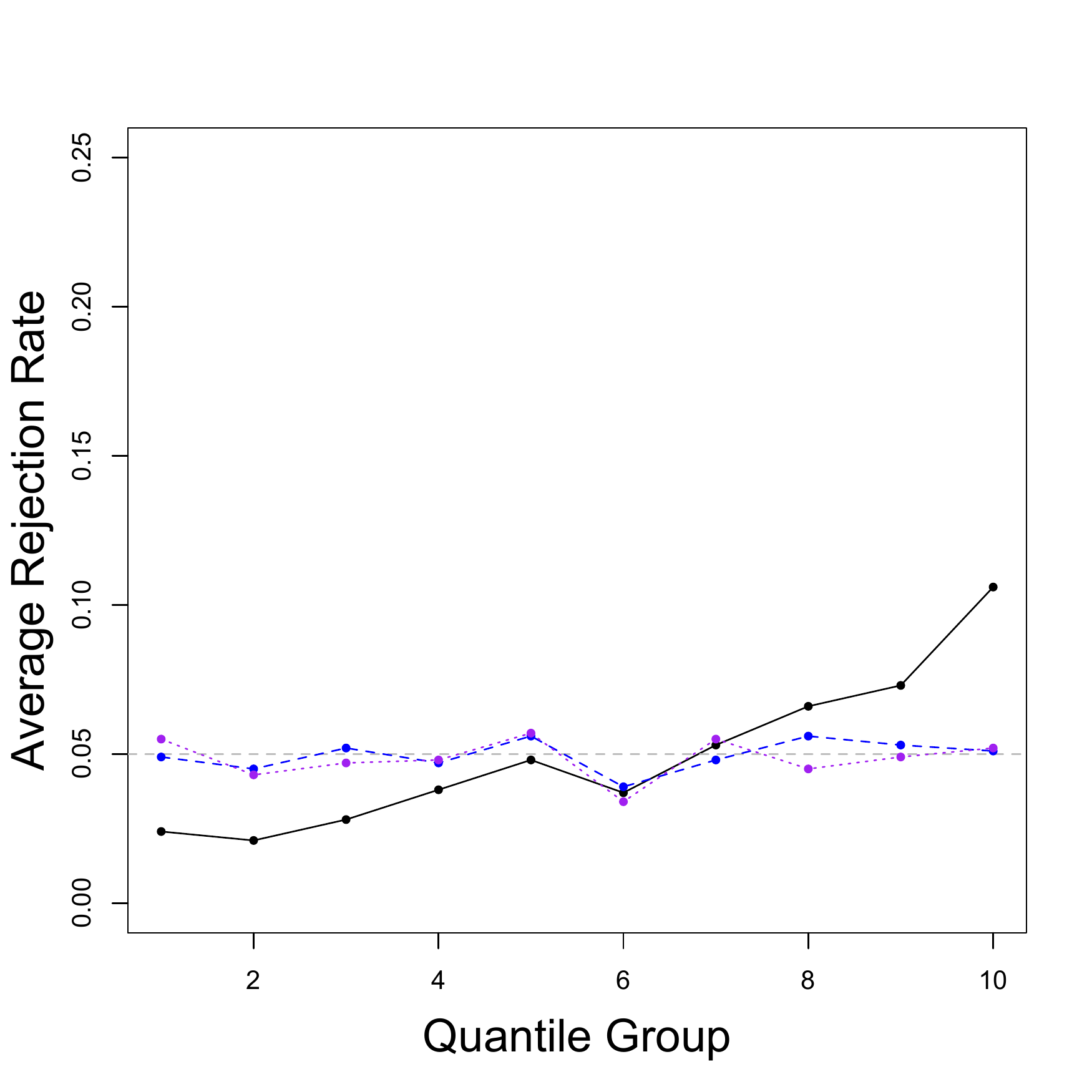}
	\caption{Moderate Correlation model, where the Mahalanobis distance is defined using $(\mathbf{X}_1, \mathbf{X}_2, \mathbf{X}_3, \mathbf{X}_4)$.}
	\label{fig:conditionalPropertiesMisspecifiedModerateLinearRawMD}
	\end{subfigure}
	~
  \begin{subfigure}[t]{0.45\textwidth}
	\centering
	\includegraphics[scale=0.4]{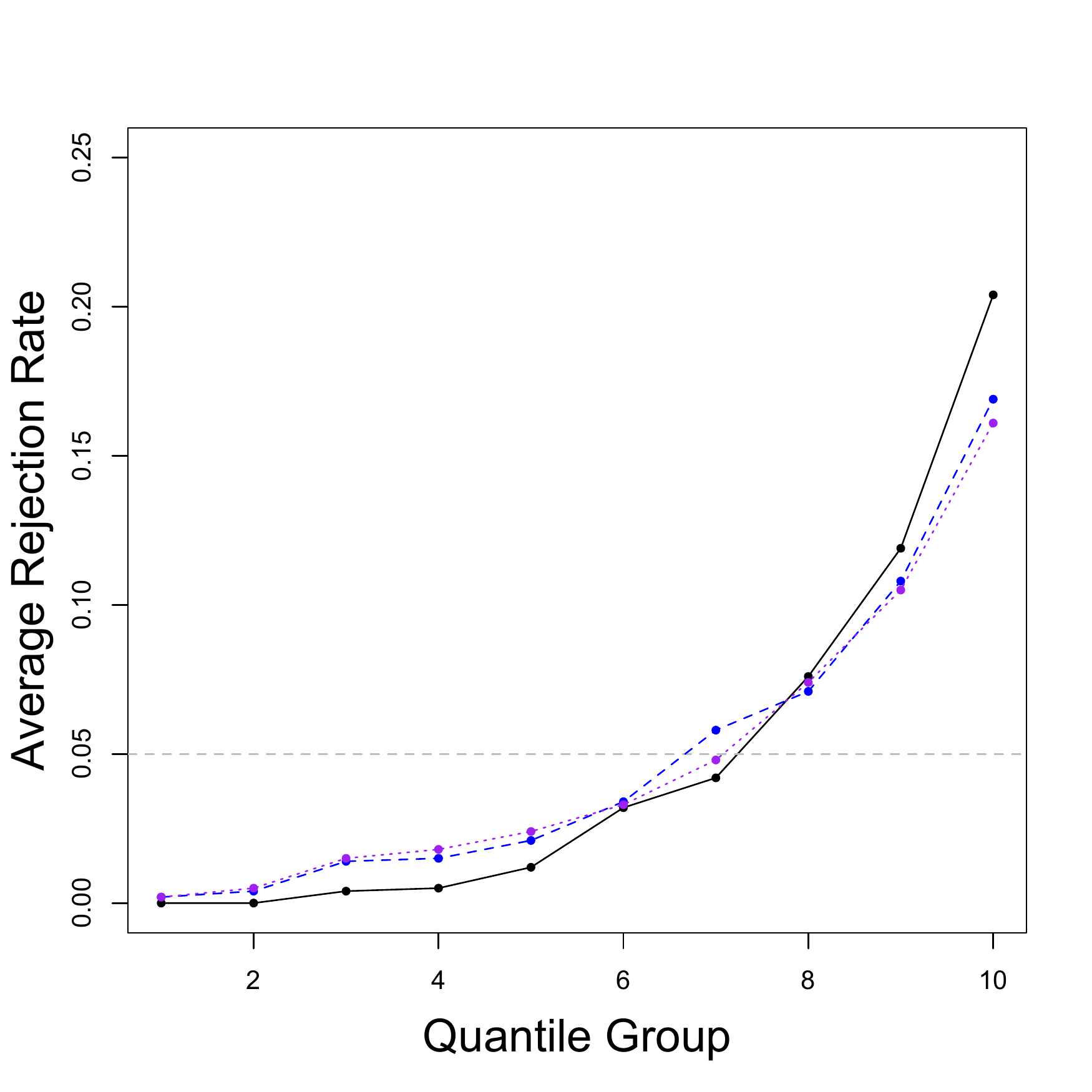}
	\caption{Moderate Correlation model, where the Mahalanobis distance is defined using $(\mathbf{X}_1^2, \mathbf{X}_2, \mathbf{X}_3^2, \mathbf{X}_4)$.}
	\label{fig:conditionalPropertiesMisspecifiedModerateLinearRealMD}
	\end{subfigure}

	\begin{subfigure}[t]{0.45\textwidth}
	\centering
	\includegraphics[scale=0.4]{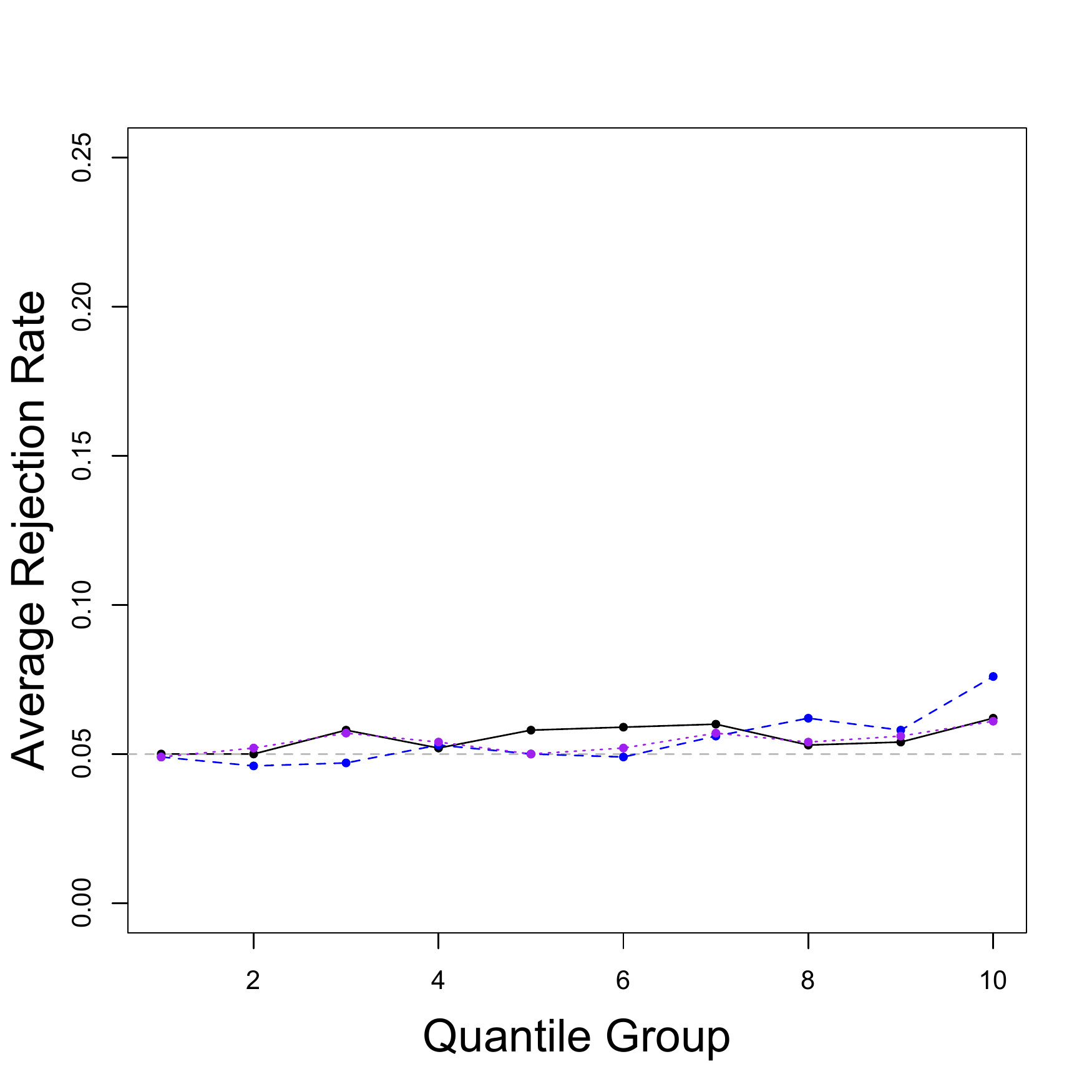}
	\caption{No Correlation model, where the Mahalanobis distance is defined using $(\mathbf{X}_1, \mathbf{X}_2, \mathbf{X}_3, \mathbf{X}_4)$.}
	\label{fig:conditionalPropertiesMisspecifiedWeakLinearRawMD}
	\end{subfigure}
	~
  \begin{subfigure}[t]{0.45\textwidth}
	\centering
	\includegraphics[scale=0.4]{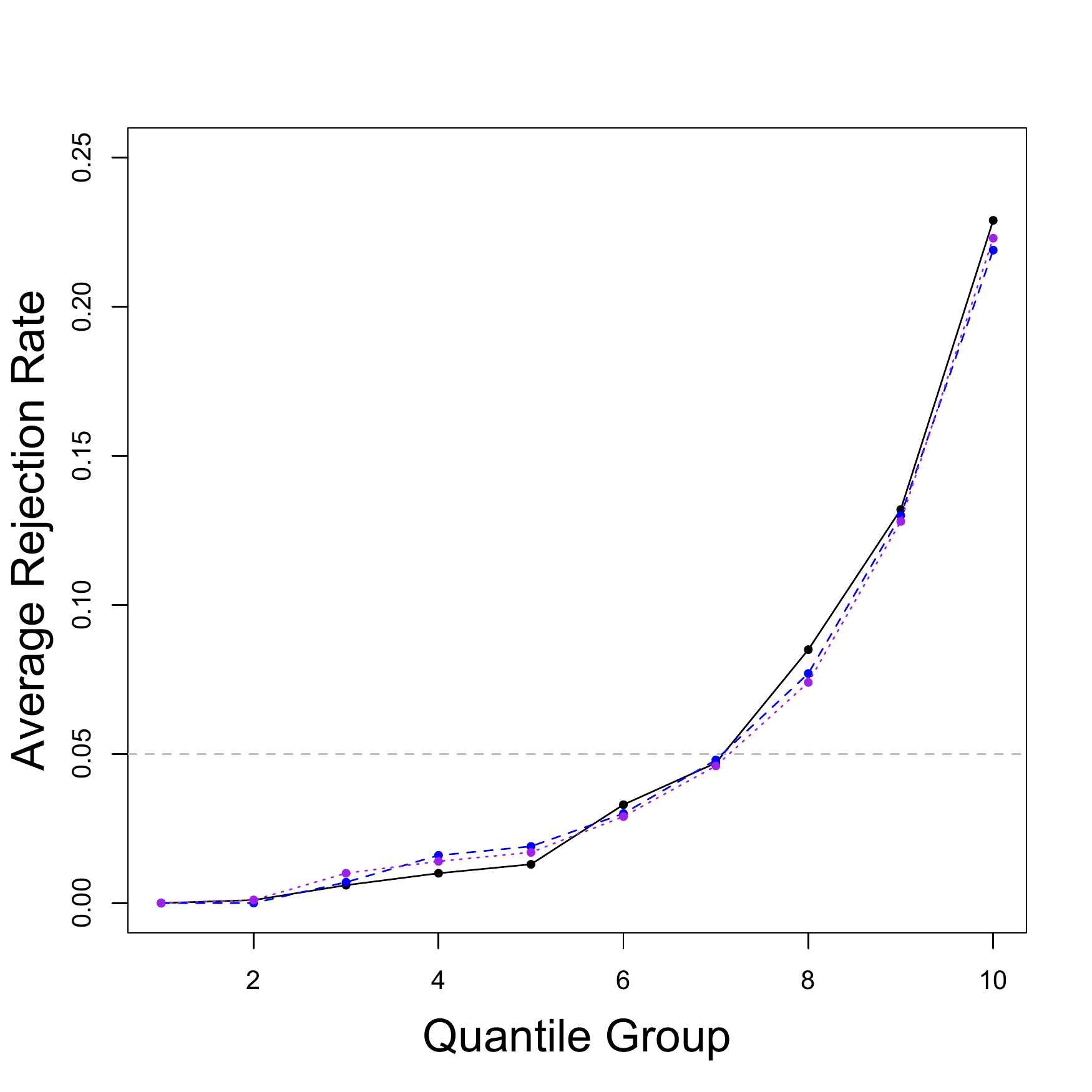}
	\caption{No Correlation model, where the Mahalanobis distance is defined using $(\sqrt{|\mathbf{X}_1|}, \mathbf{X}_2^2, \sqrt{|\mathbf{X}_3|}, \mathbf{X}_4^2)$.}
	\label{fig:conditionalPropertiesMisspecifiedWeakLinearRealMD}
	\end{subfigure}

	 \includegraphics[scale=0.5]{positiveNegativeHeterogeneousDifferentDistributionsConditionalLegend.pdf}

	\caption{The rejection rate of the randomization tests within each quantile group of the Mahalanobis distance when the potential outcomes were generated from the Moderate Correlation model (\ref{eqn:potentialOutcomesModelMisspecifiedModerateLinear}) or the No Correlation model (\ref{eqn:potentialOutcomesModelMisspecifiedWeakLinear}). In Figures \ref{fig:conditionalPropertiesMisspecifiedModerateLinearRawMD} and \ref{fig:conditionalPropertiesMisspecifiedWeakLinearRawMD}, the Mahalanobis distance is defined using the raw covariates $(\mathbf{X}_1, \mathbf{X}_2, \mathbf{X}_3, \mathbf{X}_4)$; in Figures \ref{fig:conditionalPropertiesMisspecifiedModerateLinearRealMD} and \ref{fig:conditionalPropertiesMisspecifiedWeakLinearRealMD}, the Mahalanobis distance is defined using the functions of the covariates that are linearly related with the potential outcomes for each model ($(\mathbf{X}_1^2,\mathbf{X}_2,\mathbf{X}_3^2,\mathbf{X}_4)$ and $(\sqrt{|\mathbf{X}_1|}, \mathbf{X}_2^2, \sqrt{|\mathbf{X}_3|}, \mathbf{X}_4^2)$, respectively).}
	\label{fig:conditionalPropertiesMisspecifiedModels}

\end{figure}

 Similar to Section \ref{ss:conditionalProperties}, we also examined the conditional performance of the randomization tests when the potential outcomes were generated from the Moderate Correlation and No Correlation models. Figure \ref{fig:conditionalPropertiesMisspecifiedModels} shows the rejection rate of each randomization test for each quantile group for each potential outcome model, where we followed the same quantile-binning procedure as Section \ref{ss:conditionalProperties}. In particular, in the left-hand plots of Figure \ref{fig:conditionalPropertiesMisspecifiedModels}, the Mahalanobis distance is defined \textit{using the raw covariates} $(\mathbf{X}_1, \mathbf{X}_2, \mathbf{X}_3, \mathbf{X}_4)$, whereas in the right-hand plots it is defined \textit{using the functions of the covariates that are linearly related to the potential outcomes}, i.e.,$(\mathbf{X}_1^2,\mathbf{X}_2,\mathbf{X}_3^2,\mathbf{X}_4)$ and $(\sqrt{|\mathbf{X}|}_1,\mathbf{X}_2^2,\sqrt{|\mathbf{X}|}_3,\mathbf{X}_4^2)$ for the Moderate Correlation and No Correlation models, respectively.

 When the Mahalanobis distance is defined using $(\mathbf{X}_1, \mathbf{X}_2, \mathbf{X}_3, \mathbf{X}_4)$, the results are similar to those presented in Section \ref{ss:conditionalProperties}: The unconditional randomization test using $\hat{\tau}_{int}$ and the conditional randomization test using $\hat{\tau}_{sd}$ are conditionally valid across quantile groups, while the rejection rate of the unconditional randomization test using $\hat{\tau}_{sd}$ increases with covariate imbalance. For the No Correlation model, even the unconditional randomization test using $\hat{\tau}_{sd}$ appears to be conditionally valid across quantile groups; this is because the covariates are not correlated with the outcome, and thus the treatment effect is not confounded by covariate imbalances in $(\mathbf{X}_1, \mathbf{X}_2, \mathbf{X}_3, \mathbf{X}_4)$.

 However, when the Mahalanobis distance is defined using the functions of the covariates that are linearly related to the potential outcomes, the rejection rate of all the randomization tests are monotonically increasing in the covariate imbalance defined by this Mahalanobis distance. This is because the treatment effect \textit{is confounded} by covariate imbalances in $(\mathbf{X}_1^2,\mathbf{X}_2,\mathbf{X}_3^2,\mathbf{X}_4)$ and $(\sqrt{|\mathbf{X}|}_1,\mathbf{X}_2^2,\sqrt{|\mathbf{X}|}_3,\mathbf{X}_4^2)$ for the Moderate Correlation and No Correlation models, respectively. Because none of the randomization tests incorporate these functions of the covariates, we see this monotonic behavior in the rejection rate for all randomization tests, as shown in Figures \ref{fig:conditionalPropertiesMisspecifiedModerateLinearRealMD} and \ref{fig:conditionalPropertiesMisspecifiedWeakLinearRealMD}. In other words, similar to how the unconditional randomization test using $\hat{\tau}_{sd}$ does not adjust for linear imbalances in the covariates and thus exhibited this monotonic behavior in Section \ref{s:simulations}, the conditional randomization test using $\hat{\tau}_{sd}$ and the unconditional randomization test using $\hat{\tau}_{int}$ similarly do not fully account for imbalances in $(\mathbf{X}_1^2,\mathbf{X}_2,\mathbf{X}_3^2,\mathbf{X}_4)$ or $(\sqrt{|\mathbf{X}|}_1,\mathbf{X}_2^2,\sqrt{|\mathbf{X}|}_3,\mathbf{X}_4^2)$, and thus we again see the monotonic behavior in Figures \ref{fig:conditionalPropertiesMisspecifiedModerateLinearRealMD} and \ref{fig:conditionalPropertiesMisspecifiedWeakLinearRealMD}. The conditional randomization test using $\hat{\tau}_{sd}$ and the unconditional randomization test using $\hat{\tau}_{int}$ are only accounting for imbalances in $(\mathbf{X}_1,\mathbf{X}_2,\mathbf{X}_3,\mathbf{X}_4)$. This also suggests why, in Figure \ref{fig:conditionalPropertiesMisspecifiedModerateLinearRealMD} (when the covariates are moderately correlated with the outcome), the monotonicity of the rejection rate for these two tests is less pronounced than that of the unconditional randomization test using $\hat{\tau}_{sd}$, whereas in Figure \ref{fig:conditionalPropertiesMisspecifiedWeakLinearRealMD} (when the covariates are not correlated with the outcome), the behavior of the rejection rate for all the randomization tests is essentially the same.

 In summary, when the Mahalanobis distance (or test statistic $\hat{\tau}_{int}$) is defined using functions of the covariates that are moderately correlated with the potential outcomes, then it is still beneficial to use our conditional randomization test (or the unconditional randomization test using $\hat{\tau}_{int}$) over the unconditional randomization test using $\hat{\tau}_{sd}$. Furthermore, the equivalence of the unconditional randomization test using $\hat{\tau}_{int}$ and our conditional randomization test appears to still hold when the regression used to construct $\hat{\tau}_{int}$ is misspecified. Finally, the unconditional randomization test using $\hat{\tau}_{int}$ and our conditional randomization test appear to be valid across various degrees of imbalance in functions of the covariates used to define $\hat{\tau}_{int}$ or the Mahalanobis distance. However, this does not guarantee that these tests will be conditionally valid across covariate imbalances that are not captured by $\hat{\tau}_{int}$ or the Mahalanobis distance but nonetheless confound treatment effect estimates. Regardless, both the unconditional and conditional performance of our conditional randomization test and the unconditional randomization test using $\hat{\tau}_{int}$ appear to be preferable to those of the unconditional randomization test using $\hat{\tau}_{sd}$ if covariates are correlated with outcomes, and otherwise they appear to be equivalent.

\newpage
\bibliography{conditionalRandomizationTestsBib}

\bibliographystyle{apa-good}

\end{document}